\documentclass[twocolumn,pre,superscriptaddress,floatfix,amsmath,amssymb,aps]{revtex4-2}
\pdfoutput=1

\usepackage[utf8]{inputenc}
\usepackage{natbib}
\usepackage{array}
\usepackage{booktabs}
\usepackage{graphicx}
\usepackage{paralist}
\usepackage{color,soul}
\usepackage[colorlinks=true, allcolors=blue]{hyperref}
\usepackage{mathtools}
\usepackage{amsmath}
\usepackage{dcolumn}
\usepackage{amssymb}
\usepackage{gensymb}
\usepackage{mathdesign}
\usepackage{siunitx}
\usepackage{accents}
\usepackage{bm}
\usepackage{dcolumn}

\graphicspath{{Figures/}}

\begin{document}

\title{Molecular Dynamics Simulations of Binary Sphere Mixtures}
\author{Joseph M. Monti}
\affiliation{Sandia National Laboratories, Albuquerque, NM 87185, USA}
\author{Gary S. Grest}
\affiliation{Sandia National Laboratories, Albuquerque, NM 87185, USA}
\date{\today}

\begin{abstract}
Explicit  simulations of fluid mixtures of highly size-dispersed particles are constrained by numerical challenges associated with identifying pair-interaction neighbors. 
Recent algorithmic developments have ameliorated these difficulties to an extent, permitting more efficient simulations of systems with many large and small particles of disperse sizes.
We leverage these capabilities to perform molecular dynamics simulations of binary sphere mixtures with elastically stiff particles approaching the hard sphere limit and particle size ratios of up to 50, approaching the colloidal limit.
The systems considered consist of 500 large particles and up to nearly 3.6 million small particles with total particle volume fractions up to 0.51.
Our simulations confirm qualitative predictions for correlations between large particles previously obtained analytically and for simulations employing effective depletion interactions, but also reveal additional insights into the near-contact structure that result from the explicit treatment of the small particle solvent.
No spontaneous crystal nucleation was observed during the simulations, suggesting that nucleation rates in the fluid-solid coexistence region are too small to observe crystal nucleation for feasible simulation system sizes and timescales.
\end{abstract}

\maketitle

\section{Introduction}
The binary hard sphere (BHS) mixture is a conceptually simple model that provides useful context for more complicated systems, including colloidal suspensions~\cite{lekkerkerker1992,imhof1995,zhu1997,weeks2000,royall2013} and granular packing~\cite{furnas1931,prasad2017,srivastava2021}.
Mapping the phase diagram of BHS mixtures is a classic problem that has garnered experimental~\cite{pusey1986} and analytical~\cite{biben1991,rosenfeld1994,dijkstra1999} treatments.
Notably,~\citet{dijkstra1999} provided quantitative predictions for the phase boundaries for binary mixtures for a range of large-to-small particle size ratios.
The behavior of binary mixtures in the stable fluid and stable fluid-solid coexistence region of the phase diagram is of particular interest as a prototype for colloidal suspensions.
Despite the relative simplicity of the BHS model---particles of two diameters interacting via a steep, purely repulsive potential---computational challenges have limited numerical simulations attempting to address this problem to relatively small particle size ratios.

The principal difficulties in simulating systems of particles of highly disparate sizes arises from: (1) the computational inefficiency of conventional neighboring algorithms in determining prospective interaction partners and (2) from the slow migration of large particles induced by collisions with small particles, thus requiring protracted simulation run times.
The former is mainly an issue for conventional molecular dynamics (MD) methods while the latter is endemic to all simulation techniques.

Single component systems in which the particles move in an implicit solvent can be easily simulated at essentially arbitrary volume fractions, but modeling large solute particles in an explicit solvent of even a modest number of solvent particles renders most simulation methods intractable.
A common mitigation strategy is to develop an effective depletion potential for a single pair of large particles embedded in a small particle fluid~\cite{attard1989,biben1996,dickman1997,gotzelmann1998,dijkstra1999,ashton2011} for use in MD or Monte Carlo simulations~\cite{biben1996,dickman1997,dijkstra1999}.
However, these approaches omit three particle and higher order correlations that grow in importance with increasing large particle concentration~\cite{lue1999,malherbe2001,grest2011,kobayashi2021}.
More accurate explicit simulation methods exist, including event-driven MD~\cite{lue1999,henderson2005,alawneh2008,lazaro2019,bommineni2020,pieprzyk2020,pieprzyk2021} or cluster-based algorithms~\cite{dress1995,buhot1998,lue1999,malherbe2001,malherbe2007}, but numerical studies employing these techniques have been limited to small size ratios and/or low particle volume fractions.

Recent algorithmic developments of efficient particle-size-based neighbor binning styles~\cite{ogarko2012,krijgsman2014,stratford2018,shire2021} that have been implemented into the MD package LAMMPS~\cite{thompson2022} permit simulations of unprecedented particle size ratios~\cite{srivastava2021,monti2022}.
This computational framework, which can be applied to both frictionless and frictional particles, is capable of simulating millions of particles and can feasibly reach particle size ratios of order 100 and perhaps larger for binary mixtures.
In this work, we use this capability to simulate binary mixtures of hard spheres for a range of particle volume fractions and ratios of particle diameters between large and small particles of up to 50.
The simulations use a very stiff, linearly repulsive contact model that opposes particle overlap to approximate the hard sphere limit.

Following~\citet{dijkstra1999}, we compute the radial distribution functions (RDFs) and structure factors of large particles to show that their correlations increase in magnitude with increasing small particle volume fraction.
This phenomenon is evidenced by systematic sharpening and growth of the RDF contact values and the emergence of prominent higher order RDF peaks at separations of up to two large particle diameters, showing strong qualitative agreement with the RDFs computed by~\citet{dijkstra1999}.
Similarly, the simulations demonstrate that increasing the particle size ratio at fixed small particle volume fraction also increases correlations between large particles.
Several of our simulations traverse the fluid-solid phase boundary, for which~\citet{dijkstra1999} predicted that a stable large particle fcc crystal coexists with the large particle fluid phase.
However, no spontaneous crystallization is observed in our simulations.
Rather, clusters of large particles tend to be transient, meaning that their correlations can only be computed as time-averages.
Simulations wherein the large particles are initially arranged into fcc crystallites, conversely, do show varying degrees of resistance to melting for sufficiently high $\eta_s$, supporting the notion that the fcc crystal structure is stable above the phase boundary.
We report results for a limited set of these simulations in this work.

The manuscript is organized as follows: Sec.~\ref{sec:state} briefly outlines the expected phase behavior of the mixture for the particle volume fraction and size ratio state space explored in this work; Sec.~\ref{sec:model} details the contact model and the strengths and limitations of the current approach; Sec.~\ref{sec:Results} focuses on calculations of RDFs and their contact values (Sec.~\ref{sec:rdf}) for varying small particle volume fractions and particle size ratios, and shows structural factors for the same systems (Sec.~\ref{sec:sk}). {Lastly, Sec.~\ref{sec:stability} evaluates the stability of fcc crystallites near the fluid-solid phase boundary}.

\section{Simulation details}
\label{sec:details}
\subsection{Binary mixture configuration}
\label{sec:state}
\begin{figure}
\centering
\includegraphics[width=0.4\textwidth]{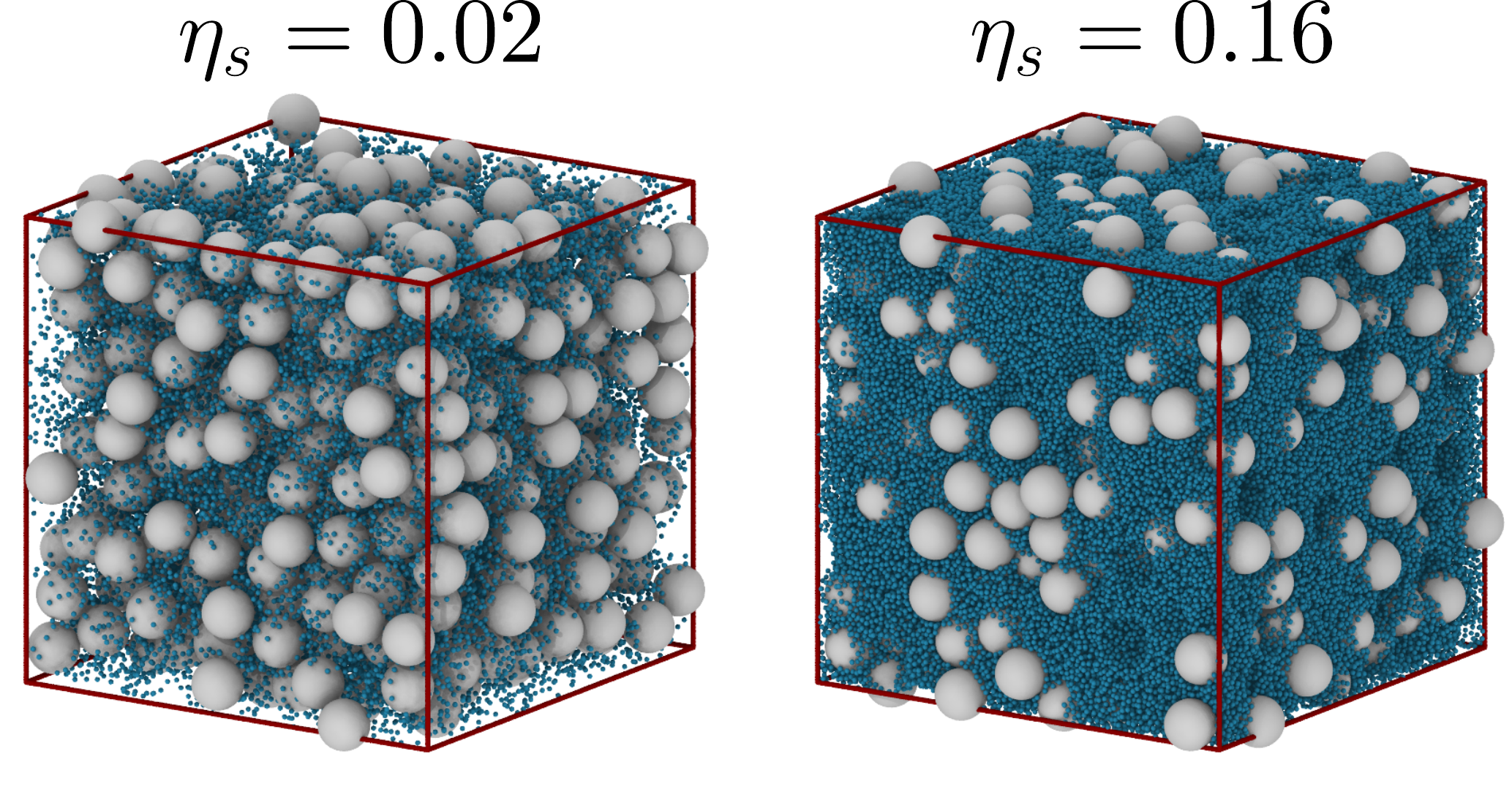}
\caption{Binary mixtures for size ratio $q = 0.1$ and large particle volume fraction $\eta_\ell = 0.35$, with small particle volume fractions $\eta_s = 0.02$ and 0.16. Images were rendered in OVITO~\cite{ovito}.}
\label{fig:fig1}
\end{figure}

The binary mixture is composed of large and small particles with diameters $\sigma_\ell$ and $\sigma_s$, respectively, with a particle size ratio $q$ denoted by $q \equiv \sigma_s/\sigma_\ell$. 
The overall particle volume fraction is $\eta = \eta_s + \eta_\ell$ in terms of the volume fractions of the individual species, $\eta_i = \pi N_i \sigma_i^3/6V$, where $i = \lbrace s,\ell\rbrace$, $N_i$ is the number of particles of each species, and $V$ is the volume of the fully periodic cubic simulation cell.
We fix $N_\ell = 500$ for all binary simulations to keep the total particle count $N = N_s + N_\ell$ tractable ($N\lesssim 3.6\times 10^6$). Where applicable, single component fluid systems are simulated with $3000$ particles to improve statistics.
Simulations are initialized by randomly placing the particles in the simulation cell without overlaps.
Figure~\ref{fig:fig1} depicts two snapshots of exemplar binary mixtures with $q = 0.1$ and with large particle volume fraction $\eta_\ell = 0.35$ for small particle volume fractions $\eta_s = 0.02$ and 0.16.

%

\begin{figure}
\centering
\includegraphics[width=0.4\textwidth]{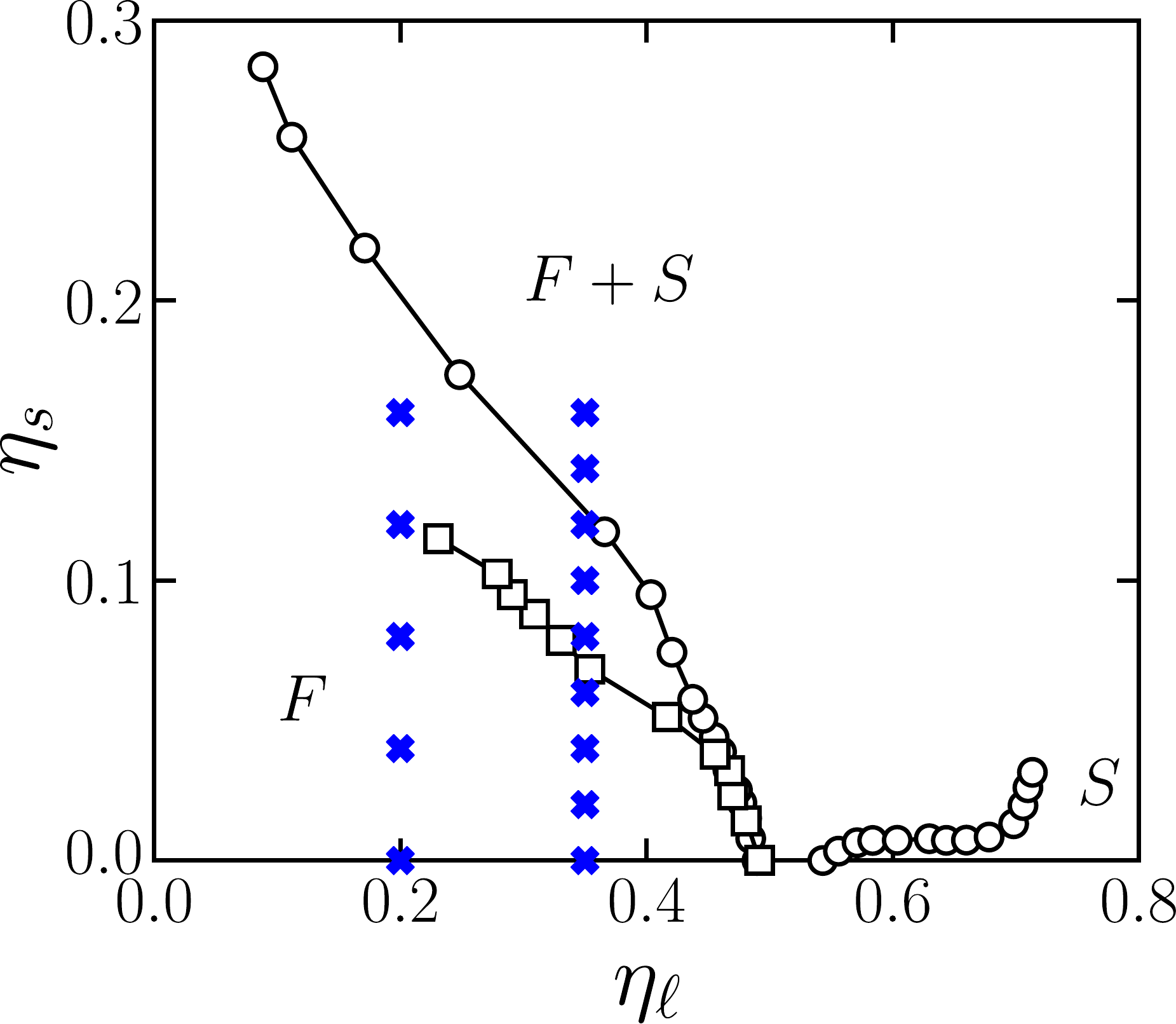}
\caption{Schematic of the binary hard sphere mixture phase diagram following~\citet{dijkstra1999}. Open symbols represent data taken from simulations reported in Ref.~\cite{dijkstra1999} (circles: $q = 0.2$, squares: $q = 0.1$) and lines are drawn to guide the eye. The fluid phase $F$, fluid-solid coexistence phase $F+S$, and solid phase $S$ are indicated. The metastable fluid-fluid $F+F$ and solid-solid $S+S$ coexistence regions~\cite{dijkstra1999} are omitted. Crosses mark the simulations performed for this work.}
\label{fig:fig2}
\end{figure}

The space of large and small particle volume fractions bounds the binary hard sphere mixture phase diagram~\cite{dijkstra1999} as shown schematically in Fig.~\ref{fig:fig2}.
Increasing the density of large particles along the $\eta_s = 0$ boundary, the hard sphere fluid has a first order melting transition, with coexisting fluid density $\eta_\ell = 0.494$ and solid density $0.545$~\cite{hoover1968}. At higher densities, the solid phase is an fcc crystal~\cite{woodcock1997,bolhuis1997}. 
As small particles are added, the density $\eta_\ell$ of large particles at the melting transition decreases and the width of the fluid-solid ($F+S$) coexistence phase increases. 
The phase boundary generally shifts to smaller $\eta_s$ at a given $\eta_\ell$ as $q$ decreases~\cite{dijkstra1999} as shown schematically in Fig.~2.
A depletion force imposed by collisions between large and small particles tends to drive large particles closer together than  in the single component fluid~\cite{attard1989,biben1996,dickman1997,gotzelmann1998,dijkstra1999,ashton2011}.
In the fluid-solid coexistence phase, these forces are predicted to be sufficiently strong to drive a portion of the large particles into a stable fcc crystal, where the proportion between large particles in the fluid and solid phases is governed by the equivalence of the chemical potential between the two phases~\cite{dijkstra1999}. 
~\citet{dijkstra1999} also showed that there are metastable fluid-fluid $F+F$ (see also~\citet{kobayashi2021}) and solid-solid $S+S$ coexistence regions. 

This work primarily considers two specific values of large particle volume fractions, $\eta_\ell = 0.2\ \text{and}\ 0.35$, for a range of values of $q$ with a minimum value of 0.02.
Note that these state points are far removed from the single component freezing point and from the metastable $F+F$ and $S+S$ coexistence regions.
To explore changes in particle correlations near the phase boundary described above, $\eta_s$ is increased systematically from zero up to $\eta_s = 0.16$ in discrete steps.
{From~\citet{dijkstra1999}, for $q = \lbrace 0.2, 0.1, 0.05\rbrace$ the phase boundary crosses $\eta_\ell = 0.35$ at $\eta_s \approx \lbrace 0.12,0.07,0.05\rbrace$ and $\eta_\ell = 0.2$ at $\eta_s \approx \lbrace 0.20,0.12, 0.07\rbrace$ (c.f. Fig. 15 of~\citet{dijkstra1999}).}

Although several of the densest overall systems considered are nominally in the fluid-solid coexistence region, we did not find evidence of large particle crystal formation~\cite{lue1999}. 
Crystal nucleation rates from the mixed fluid are expected to be minuscule for the range of $\eta$ considered~\cite{auer2001,filion2010,bommineni2020}.
{To circumvent challenges associated with nucleation kinetics, we performed additional test simulations initialized by arranging all $N_\ell = 500$ large particles into a $5\times5\times5$ unit cell fcc crystallite.
The crystallites were constructed at the highest fcc density and small particles were permitted to interpenetrate the lattice.
Notionally, systems corresponding to state points in the fluid region of the phase diagram should undergo complete melting, while systems in the coexistence region should retain at least part of the crystallite that is in equilibrium with the fluid.
The Ackland-Jones technique~\cite{ackland2006} implemented in OVITO~\cite{ovito} was used to identify large particles that remained specifically in the fcc structure over simulation time.
Results for these test simulations are shown for increasing $\eta_s$ in Sec.~\ref{sec:stability}.
}

\subsection{Contact model}
\label{sec:model}
The simulations of noncohesive particles in the hard sphere limit are conducted using the LAMMPS~\cite{thompson2022} MD simulation package.
An efficient particle-size-based neighbor binning algorithm~\cite{ogarko2012,krijgsman2014,stratford2018,shire2021} permits the simulation of large maximum particle size ratios~\cite{srivastava2021,monti2022}.
Particles collide elastically via a very stiff Hookean interaction that limits overlap and vanishes when the particles are separated by a distance $r>\sigma_{ij} = (\sigma_i+\sigma_j)/2$, where the $i$ and $j$ subscripts label the particle diameters.
Throughout this work, a single subscript is used in cases where the interacting particles have identical diameter. 
The normal force between particles with center-to-center separation $r_{ij}<\sigma_{ij}$ is $\mathbf{F}_{ij} = -K\delta_{ij}\mathbf{n}_{ij}$, where $K$ is the spring constant, $\delta_{ij} = r_{ij}-\sigma_{ij}$ is the particle overlap, and $\mathbf{n}_{ij}$ is the unit vector connecting the particle centers. To approach the hard sphere limit, a large value of $K$ is used to minimize the particle overlap.
Particles are assumed to have unit mass density so that small particles have mass $M_s = \pi/6$ and large particles have mass $M_\ell = M_s\sigma_\ell^3/\sigma_s^3$.

Simulations are conducted at dimensionless temperature $T$ and particle motion is thermally driven by coupling the particles to a Langevin thermostat~\cite{grest1986}.
For finite spring stiffness, typical particle overlaps can be estimated via the equipartition theorem as $\langle \delta_{ij}^2\rangle = T/K$.
This expression indicates that overlaps can be mitigated by increasing the spring stiffness $K$ or by reducing the temperature $T$. 
However, the simulation time step must be reduced as $K$ increases to resolve small-small particle collisions, which have a duration of $\tau_{\rm c} = \pi\sqrt{M_s/2K}$.
Thus, every $10$  fold increase in $K$ increases the required CPU-time per simulation by a factor $\sim 3$ if all other parameters are held fixed.
{Hydrodynamic interactions are omitted from this work, but could be included by following the approach of~\citet{wang2015}, for instance.}

\begin{figure}
\centering
\includegraphics[width=0.4\textwidth]{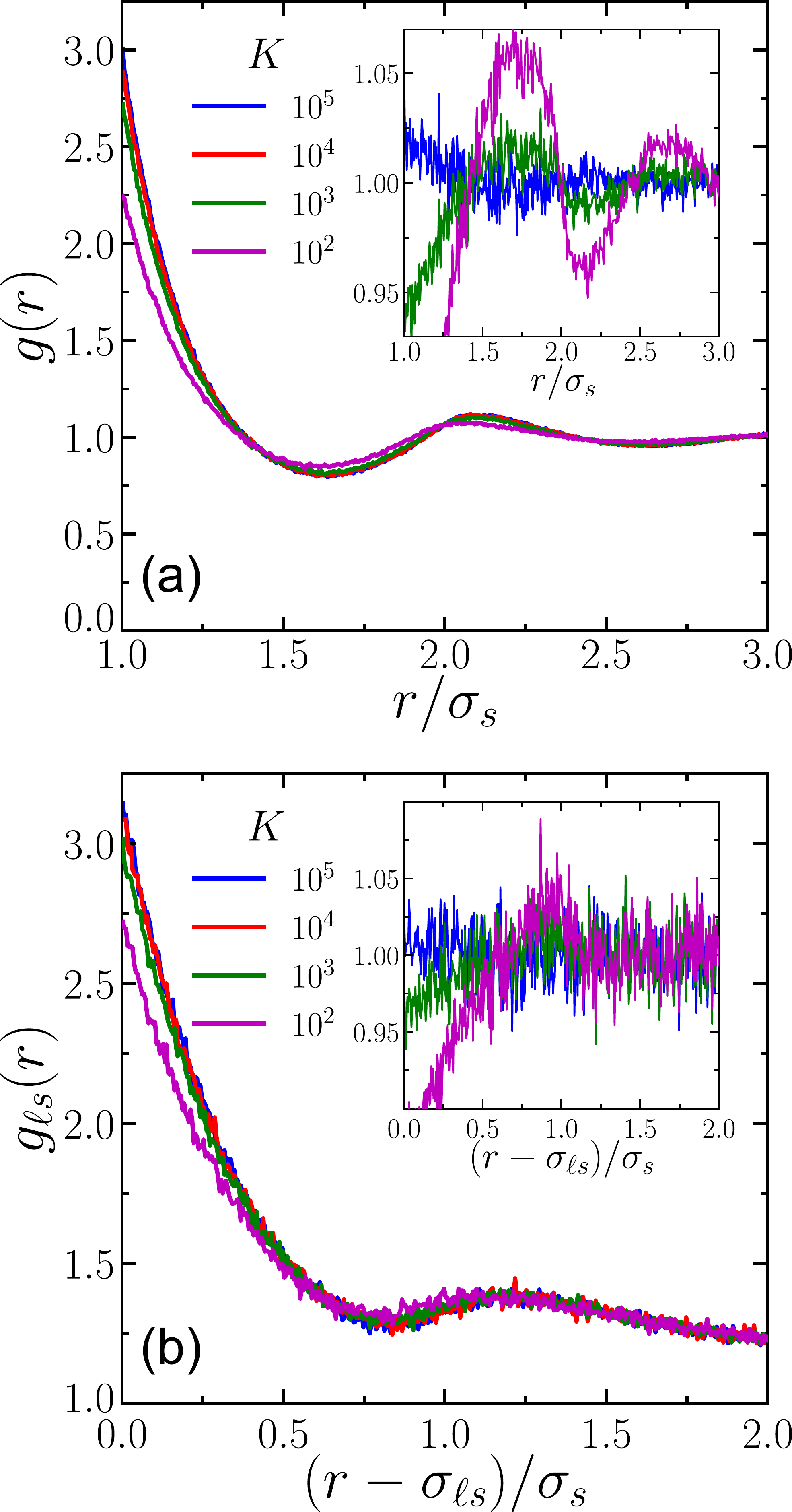}
\caption{{(a) RDFs for a single component system with $\eta = \eta_\ell = 0.35$ with varying contact stiffness $K$. (b) Inter-species RDFs $g_{\ell s}(r)$ obtained for systems with $\eta_{\ell} = 0.35$ and $\eta_{s} = 0.1$ for $q = 0.1$ using the same range of $K$ as in (a). Note that the y-axis is shifted above zero in (b). Insets: ratios of $g(r)$ (a) and $g_{\ell s}(r)$ (b) for $K \neq 10^4$ with respect to the curves for $K = 10^4$. The bin size used to compute the RDFs is $\delta r = 0.005\sigma_s$.}}
\label{fig:fig3}
\end{figure}

\begin{figure}
\centering
\includegraphics[width=0.4\textwidth]{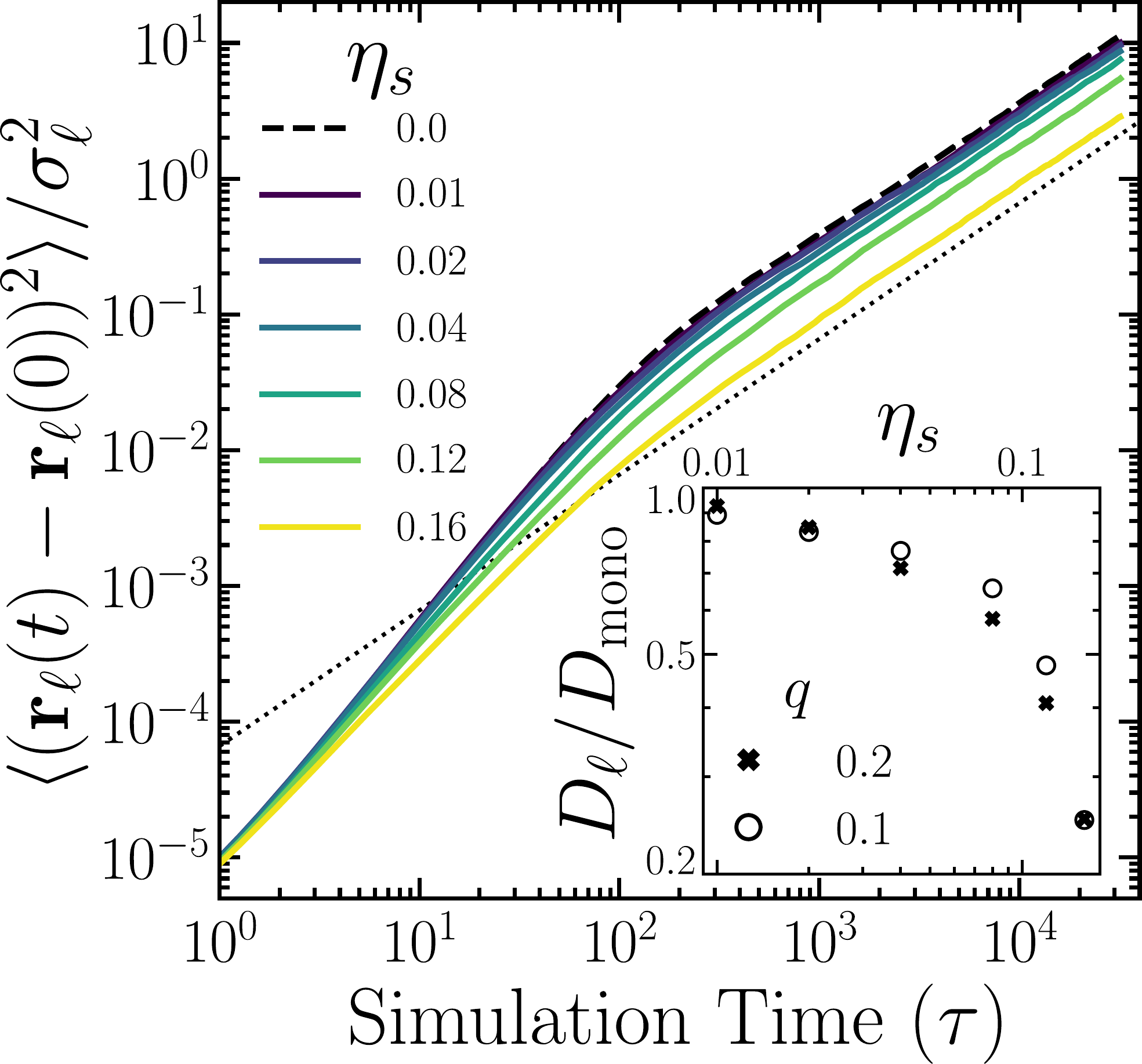}
\caption{Normalized mean squared displacement of large particles $\langle(\mathbf{r}_{\ell}(t)-\mathbf{r}_{\ell}(0))^2\rangle$ as a function of time $t$ for systems with $\eta_{\ell} = 0.35$ and $q = 0.1$, averaged over six separate reference states. The dotted black line denotes linear scaling with simulation time. Inset: ratio of the large particle diffusion constant $D_{\ell}$ to the single component value for the indicated $q$ on double logarithmic axes. Errors in the estimated values of $D_\ell$ are comparable to the symbol size.}
\label{fig:fig4}
\end{figure}

{To quantify the approach of the finite stiffness contact model to the hard sphere limit at fixed $T = 0.1$, Fig.~\ref{fig:fig3} shows comparisons of RDFs computed for $\eta_\ell = 0.35$ for increasing values of $K$ calculated over similar simulation times: for the single component RDF $g(r)$ in panel (a) and inter-species RDF $g_{\ell s}(r)$ for $q = 0.1$ and $\eta_s = 0.1$ in panel (b).}
{The RDFs are essentially independent of $K$ beyond three particle diameters in the single component case and two small particle diameters outside of contact in the inter-species case, but differences emerge near contact}.
{Both panels of Figure~\ref{fig:fig3} indicate that the contact values $g(\sigma_s)$ and $g_{\ell s}(\sigma_{\ell s})$ increase with $K$, with progressively smaller changes as $K$ rises.}
{The insets show the ratios of $g(r)$ and $g_{\ell s}(r)$ computed for $K \neq 10^4$ to those of $K = 10^4$; plotted in this way, $\lesssim10\%$ oscillations in correlations are evident for $K<10^4$, while for $K = 10^5$ the contact values increase $\lesssim 2\%$ compared to $K = 10^4$.}
Consequently, in this work we adopt the value $K = 10^4$ and $T = 0.1$, resulting in maximum overlap of $\delta_{ij} \sim 0.003\sigma_s$.
These values provide an adequate compromise between the constraints of small particle overlaps and simulating diffusive timescales for representative numbers of particles, for systems with $q$ beyond those that have been numerically tractable previously.
{Note that the similarity between $K$-dependencies for the single component and inter-species RDFs originates from the size independence of the particle interactions.}
{Particle interactions that account for particle size, e.g., the Hertz contact model, are expected to exhibit dependence of the contact values on $q$.}

The simulation time step is $\Delta t = 0.00032\tau \approx 0.02\tau_{\rm c}$ in terms of the time unit $\tau$ and small-small collision duration $\tau_{\rm c}$.
Temperature is imposed using a Langevin thermostat with a damping time value of $100\tau\sim 6200\tau_{\rm c}$.
Unless otherwise noted, all simulations are performed for a simulation time of $3.2\times 10^5\tau \approx 2\times 10^7\tau_{\rm c}$.
Radial distribution function  and structure factor calculations were performed by averaging over simulation frames separated by 0.5\% of the simulation run time, omitting the first 5\% of the run time to allow the initial configuration to equilibrate.

To illustrate large particle diffusive behavior, Fig.~\ref{fig:fig4} shows the large particle mean squared displacement $\langle(\mathbf{r}_{\ell}(t)-\mathbf{r}_{\ell}(0))^2\rangle$, averaged over all large particles, as a function of time $t$ for $q = 0.1$ and increasing $\eta_s$.
The elapsed simulation time depicted in Fig.~\ref{fig:fig4} represents 10\% of the overall simulation time.
Mean squared displacement is linear in time in the diffusive regime, and Fig.~\ref{fig:fig4} shows that large particles are able to diffuse many times their diameters in the allotted time.
The large particle diffusion coefficient $D_\ell$ decreases with increasing $\eta_s$, resulting in a downwards shift on the double logarithmic axes~\cite{imhof1995,lazaro2019}.
The inset shows the computed values of $D_\ell$ normalized by the equivalent single component result for $q = 0.2$ and $0.1$. 
These results are not meant to be exhaustive and are included here solely to demonstrate that our simulations are able to reach timescales such that computed $g(r)$ and structure factors $S(k)$ are representative of equilibrated fluid mixtures.
However, we note that the simulation times employed in this study are not sufficient to reach $\langle(\mathbf{r}_{\ell}(t)-\mathbf{r}_{\ell}(0))^2\rangle > \sigma^2_\ell$ for $q < 0.05$, and we include the corresponding data in Sec.~\ref{sec:Results} primarily to compare with larger $q$ results.

\begin{figure*}
\centering
\includegraphics[width=\textwidth]{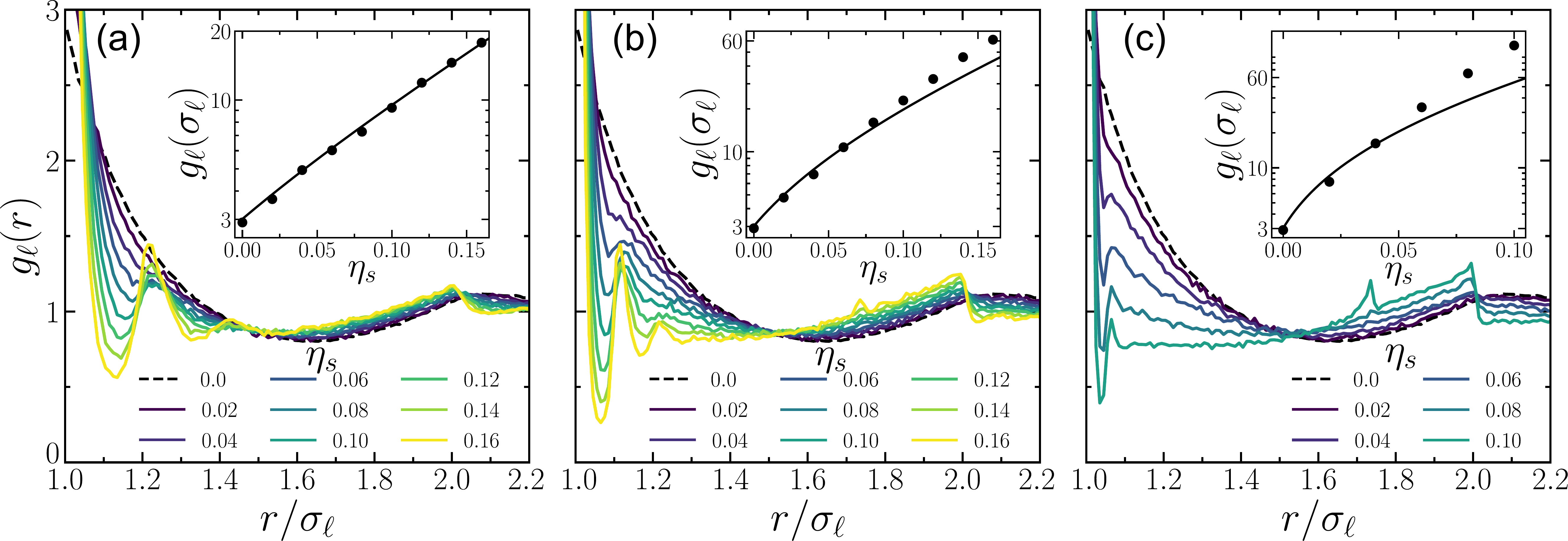}
\caption{Large particle RDFs for $\eta_{\ell} = 0.35$ for the indicated values of $\eta_{s}$ for (a) $q = 0.2\ (\sigma_{\ell} = 5)$, (b) $q = 0.1\ (\sigma_{\ell} = 10)$, and (c) $q = 0.05\ (\sigma_{\ell} = 20)$. Insets: Estimated contact values for the same data (symbols) and the corresponding predictions (lines) of~\citet{viduna2002a,viduna2002b}. Error bars for the simulation data are comparable to or smaller than the symbol size.
}
\label{fig:fig5}
\end{figure*}

\begin{figure*}
\centering
\includegraphics[width=\textwidth]{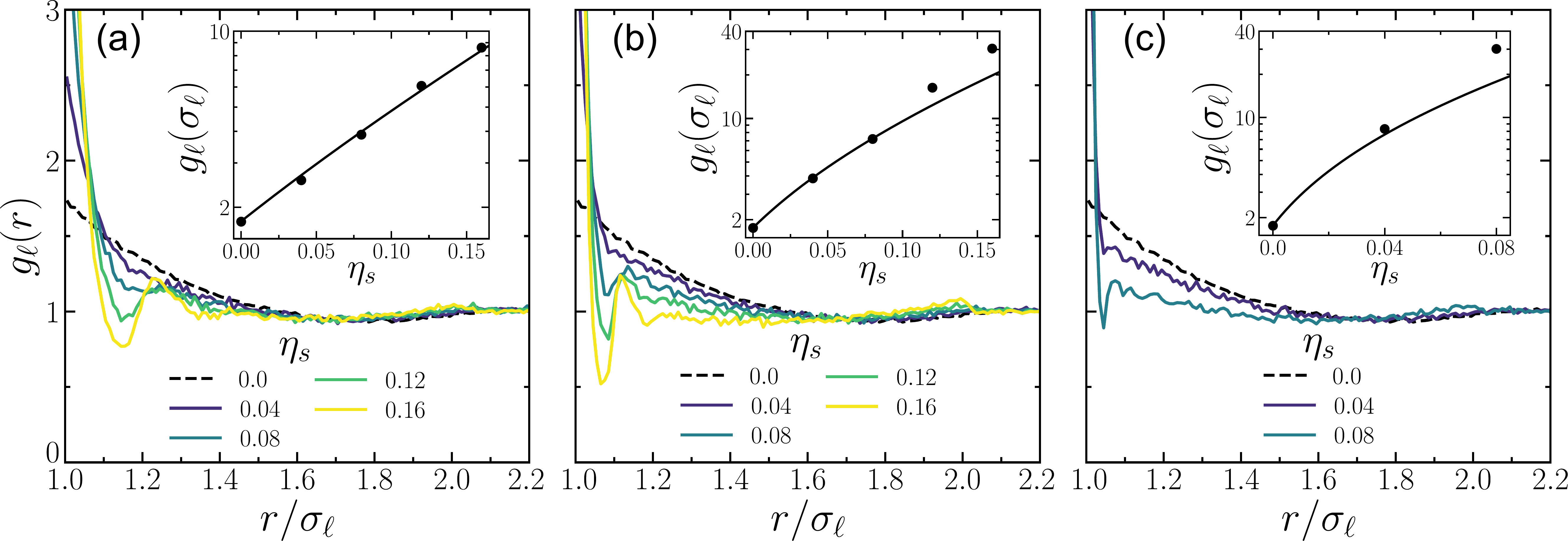}
\caption{Large particle RDFs for $\eta_{\ell} = 0.20$ for the indicated $\eta_{s}$ for (a) $q = 0.2$, (b) $q = 0.1$, and (c) $q = 0.05$. Insets: Estimated contact values for the same data (symbols) and the corresponding predictions (lines) of~\citet{viduna2002a,viduna2002b}. Error bars for the simulation data are comparable to or smaller than the symbol size.
}
\label{fig:fig6}
\end{figure*}

\section{Results}
\label{sec:Results}
\subsection{Radial distribution functions}
\label{sec:rdf}
For binary mixtures, three distinct RDFs can be defined: $g_s(r)$ and $g_\ell(r)$ for small-small and large-large pair correlations and $g_{\ell s}(r)$ for inter-species correlations.
Mainly, $g_\ell(r)$ is the quantity of interest, and indeed, in systems employing an effective potential between large particles, $g_\ell(r)$ is the only RDF available.
In this section, we first present calculations of $g_\ell(r)$ obtained via simulations of varying particle size ratios $q$ and small particle volume fractions $\eta_s$.
Unless otherwise noted, the bin size $\delta r$ used to calculate the RDFs satisfies $q\delta r = 0.01$.
The section is concluded with a brief discussion of $g_{\ell s}(r)$  for $q = 0.2$ and $0.1$.

Figure~\ref{fig:fig5} shows $g_\ell(r)$ for $\eta_\ell = 0.35$ and $q = 0.2, 0.1$, and $0.05$ with $\eta_s$ varied so that $0.35 \leq \eta \leq 0.51$.
For the single component fluid, the primary maximum at contact decays monotonically over the range $r = \sigma_\ell$ to $r \approx 1.6\sigma_\ell$.
As $\eta_s$ increases, pronounced changes in large particle spatial correlations emerge within the first $\sim 2\sigma_\ell$ compared to the single component fluid result.
Excepting the several smallest $\eta_s$ for each $q$, $g_\ell(r)$ forms oscillatory features with peaks arising near spacings corresponding to $1\times$ and $2\times$ (for the highest $\eta_s$) the small particle diameter.
These peaks are associated with configurations in which small particles are trapped in between large particles, thereby preventing their direct contact.
The effective depletion force between large particles similarly switches between attraction and repulsion as a result of the layering behavior~\cite{attard1989,dickman1997,dijkstra1999}.
Note that for smaller $q$, the first oscillation may be indistinguishable from the primary maximum, as it is located near $r/\sigma_\ell = 1+q$.
Both the primary contact peak and the secondary peak located at $r = 2\sigma_\ell$ sharpen with increasing $\eta_s$.
Further, for each of the highest two $\eta_s$ values for $q = 0.1$ and 0.05, which are nominally in the fluid-solid coexistence region, an additional peak  at $r = \sqrt{3}\sigma_\ell$ appears, indicating the formation of trigonal bi-pyramid configurations~\cite{biben1996,dijkstra1999}.
This peak is absent from results reported by~\citet{dijkstra1999} for simulations employing an effective large-large particle depletion interaction at volume fractions that similarly traversed the phase boundary, while the primary and secondary peaks in their work displayed the same trends shown here.
Over the range $1.6\lesssim r/\sigma_\ell\lesssim 2.1$ the highest $\eta$ RDFs shown in Fig.~\ref{fig:fig5} are reminiscent of those obtained for random jammed packings of monodisperse spheres (c.f.~\cite{hermes2010}, for example), implying that temporary clusters of large particles are forming.
Visual inspection of the $q = 0.05,\eta_s = 0.10$ system did not reveal persistent clusters in \textit{immediate} contact or evidence of crystallization, but cluster analysis with a cutoff distance criterion outside of contact (in units of $\sigma_s$) revealed that the fraction of large particles contained in the largest cluster varied between $0.5-0.75$ with a cutoff of 0.05 and was greater than 0.9 with a cutoff of 0.1~\cite{biben1996}.

Of particular interest are  the values of the RDF at contact $g_{ij}(\sigma_{ij})$ as they are related to the system pressure and corresponding BHS equations of state.
For finite stiffness Hookean interactions, the strength of the first neighbor peak is reduced over a width of order the particle overlap.
To estimate the contact values, we used a linear extrapolation of the RDF values just outside of contact.
For the purposes of these calculations, $g_{ij}(\sigma_{ij})$ was computed using a $q$-independent bin size of $\delta r = 0.005\sigma_s$; the extrapolation was based on a linear fit to the first four bins (a range of $0.02\sigma_s$) outside of contact. 
Similar contact values were obtained by integrating over the range of RDF values corresponding to overlap.
To quantify the error on the contact value estimates, the same extrapolation procedure was performed on a subset of the simulation frames used to compute the overall RDF: for $g_\ell(\sigma_\ell)$ three subsets of equal number were computed and four subsets for $g_{\ell s}(\sigma_{\ell s})$.
In cases where the range of extrapolated contact values across the subsets is larger than the symbols, error bars are drawn to indicate that range.

Contact value results from earlier simulation studies at the volume fractions considered here are largely absent from the literature.
Nevertheless, we simulated a single special system with $q = 0.1$ and $\eta_\ell = \eta_s = 0.1$ to compare with the contact value obtained by~\citet{malherbe2007}, who used a selective-pivot sampling algorithm with $N_\ell = 66$ large spheres.
They found $g_\ell(\sigma_\ell) = 7.1 \pm 0.1$ for this system.
Using the procedure outlined above and with $N_\ell = 500$, we obtained $g_\ell(\sigma_\ell) \approx 7.7$ representing an approximately $8\%$ increase over the~\citet{malherbe2007} result, with a range of contact values given by $7.6 \leq g_\ell(\sigma_\ell) \leq 8.1$ across the three subsets.
Note that we simulated this system for double the simulation time of our other simulations to mitigate the longer equilibration times inherent to more dilute mixtures.

\begin{figure}
\centering
\includegraphics[width=0.4\textwidth]{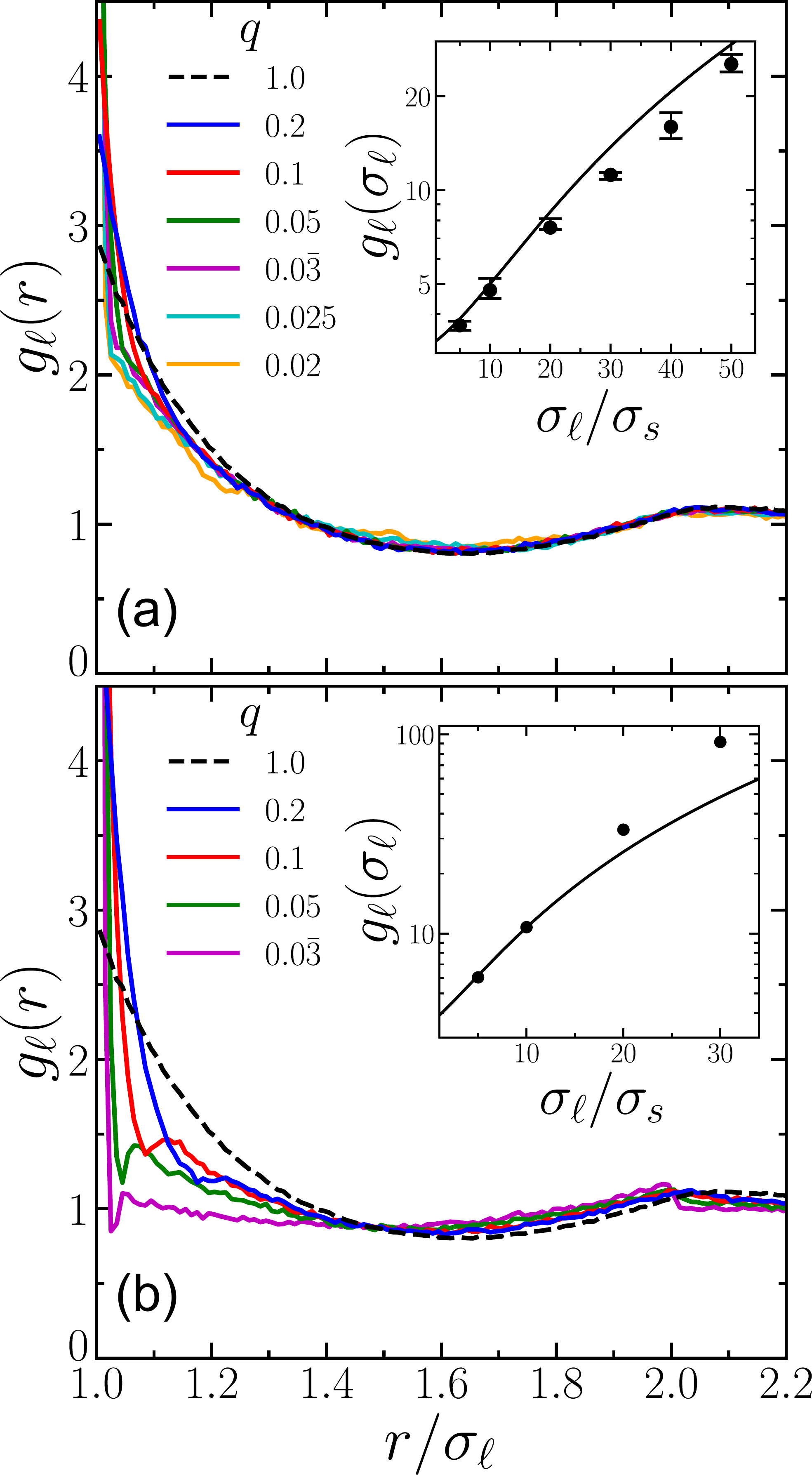}
\caption{Large particle RDFs $g_{\ell}(r)$ for $\eta_{\ell} = 0.35$ for the indicated values of $q$ for (a) $\eta_{s} = 0.02$ and (b) $\eta_{s} = 0.06$. The single component fluid results ($q = 1.0$) are included to provide a basis of comparison for the RDFs of smaller $q$, but have no direct correspondence with $\eta_s > 0$ data. Insets: Estimated contact values for the same data (symbols) and the corresponding of predictions (lines) of~\citet{viduna2002a,viduna2002b}. Error bars in (b) are comparable to or smaller than the symbol size.
}
\label{fig:fig7}
\end{figure}

\begin{figure}
\centering
\includegraphics[width=0.4\textwidth]{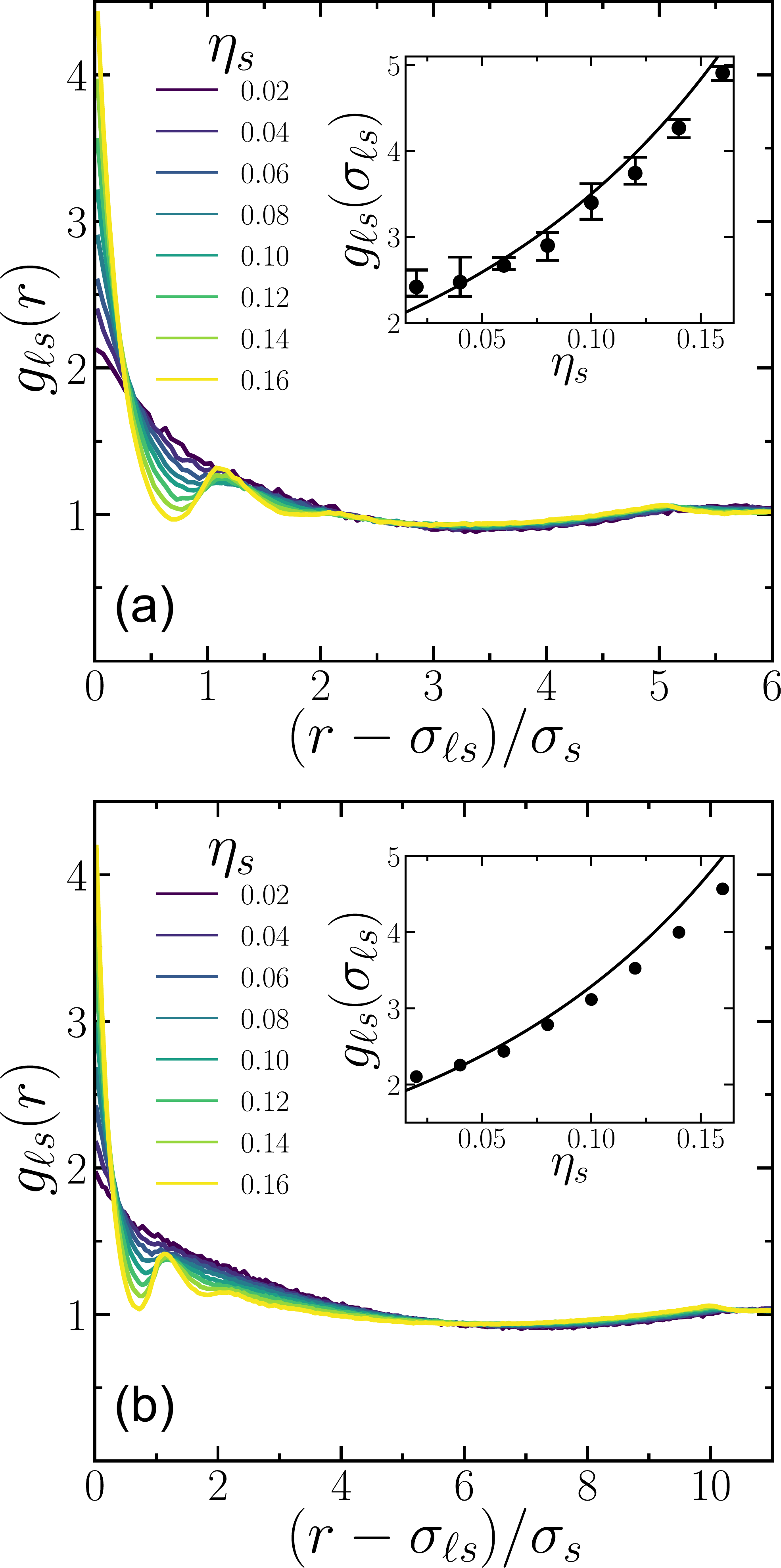}
\caption{Inter-species RDFs $g_{\ell s}(r)$ obtained for systems with $\eta_{\ell} = 0.35$ for the indicated $\eta_{s}$ for (a) $q = 0.2$ and (b) $q = 0.1$. Insets: Estimated contact values for the same data (symbols) and the corresponding  predictions (lines) of~\citet{viduna2002a,viduna2002b}. Error bars in (b) are comparable to or smaller than the symbol size.}
\label{fig:fig8}
\end{figure}

{For the single component system, contact values can be estimated via the empirical expression due to~\citet{kolafa2004}, which evaluates to $g(\sigma_s) \approx 3.01$ for $\eta = 0.35$.}
We obtained $g(\sigma_s) \approx 2.91$, or 3\% smaller for  $\eta_s = 0$ as shown in Fig.~\ref{fig:fig5}.
Various other empirical contact value formulae have been derived for the BHS mixture~\cite{henderson2000,viduna2002a,viduna2002b,alawneh2008,santos2009}, usually building on the Boubl\'{i}k–Mansoori–Carnahan–Starling–Leland (BMCSL) equation of state~\cite{boublik1970,mansoori1971}.
We found that expressions due to~\citet{viduna2002a,viduna2002b} (VS) best fit our data over the expected domain of validity, i.e., relatively high values of $q$ and of $N_\ell/N$.
Written as an expansion in terms of powers of the particle diameters, the VS contact values are~\cite{viduna2002b}
\begin{equation}
\begin{split}
    g_{ij}(\sigma_{ij}) = \frac{1}{1-\eta} + \eta\frac{\left(3-\eta+\eta^2/2 \right)}{2\left(1-\eta \right)^2}\frac{\xi_2}{\xi_3}\left(\frac{\sigma_i\sigma_j}{\sigma_{ij}}\right) \\
    + \eta^2\frac{\left(2-\eta-\eta^2/2 \right)}{2\left(1-\eta \right)^3}\frac{2\xi_2^2+\xi_1\xi_3}{3\xi_3^2}\left(\frac{\sigma_i\sigma_j}{\sigma_{ij}}\right)^2,
    \end{split}
\end{equation}
where $\xi_m = (N_s\sigma_s^m+N_\ell\sigma_\ell^m)/N$ are moments of the particle size distribution.
We also tested the expressions given by~\citet{alawneh2008} and by~\citet{santos2009} but found these showed poorer overall agreement with our data.
The VS formula is plotted along with contact values estimated from our data in the insets of each RDF figure.
Note that the inset axes are mainly semi-logarithmic, as the contact value grows nearly exponentially with both $\eta_s$ and $\sigma_\ell$~\cite{roth2000,alawneh2008,santos2009}.
By construction, the VS formula (and others) are accurate for the single component system.
The accuracy decreases with increasing particle size disparity and as one approaches the fluid-solid coexistence region.
This behavior is clear from the insets of Fig.~\ref{fig:fig5}: for $q = 0.2$, the VS formula goes through the data points over the entire range of $\eta_s$ considered; however, as $q$ decreases to 0.1 and below, the VS formula increasingly deviates from the data.
Since increasing particle stiffness slightly increases the contact value (e.g., see Fig.~\ref{fig:fig3}) the discrepancy cannot be accounted for by the finite compliance of the simulated particles.

\begin{figure*}
\centering
\includegraphics[width=\textwidth]{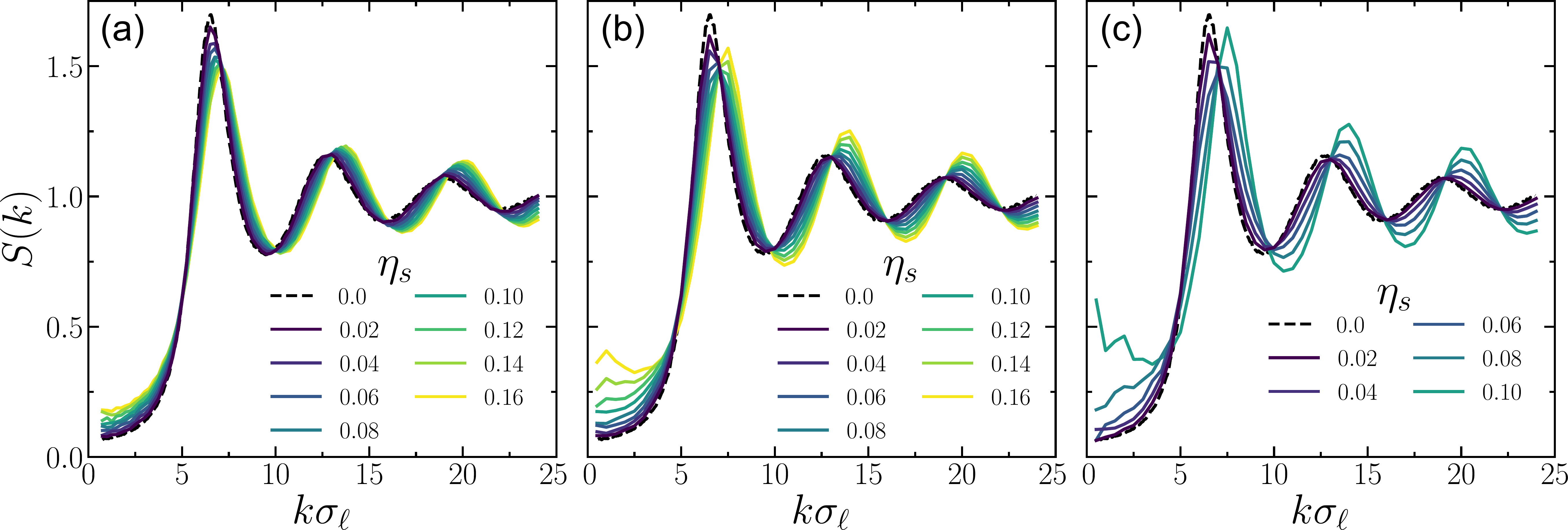}
\caption{Structure factor $S(k)$ for $\eta_{\ell} = 0.35$ for the indicated $\eta_{s}$ for (a) $q = 0.2$, (b) $q = 0.1$, and  (c) $q = 0.05$.}
\label{fig:fig9}
\end{figure*}
\begin{figure*}
\centering
\includegraphics[width=\textwidth]{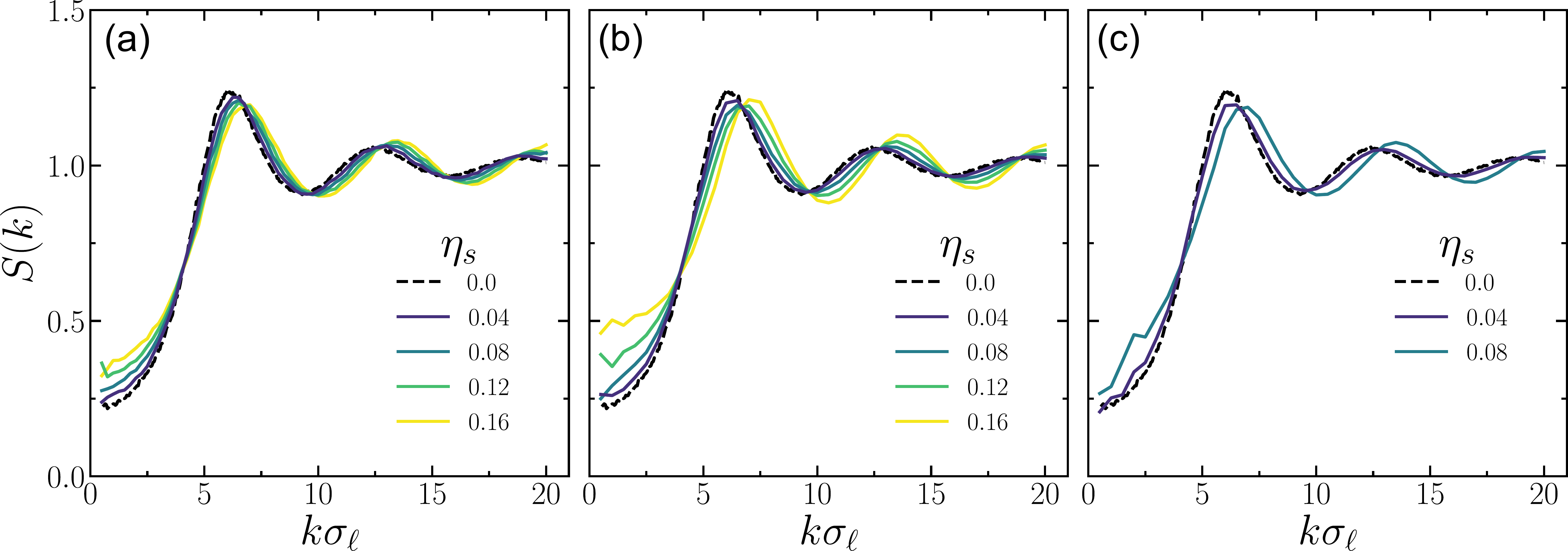}
\caption{Structure factor $S(k)$ for $\eta_{\ell} = 0.20$ for the indicated $\eta_{s}$ for (a) $q = 0.2$, (b) $q = 0.1$, and (c) $q = 0.05$.}
\label{fig:fig10}
\end{figure*}

To contrast with the results shown above, results for $\eta_\ell = 0.20$ and the same range of $q$ and $\eta_s$ are shown in Fig.~\ref{fig:fig6}.
With the exceptions of the highest single $\eta_s$ value simulated for $q = 0.1$ and 0.05, each of these systems are contained within the fluid region of the BHS mixture phase diagram {or lie directly on its border}.
The results in Fig.~\ref{fig:fig6} exhibit many of the same qualitative features and trends as were apparent for $\eta_\ell = 0.35$ in Fig.~\ref{fig:fig5}.
Notably, the primary peak and the secondary peak at $r = 2\sigma_\ell$ once again sharpen with increasing $\eta_s$, and the oscillatory behavior associated with trapped small particles also emerges.
Only the first oscillation near $r/\sigma_\ell = 1+q$ is apparent in Fig.~\ref{fig:fig6} at $\eta_\ell = 0.20$.
{For the single component fluid, we obtained a contact value of 1.75, a less than 1\% deviation from the~\citet{kolafa2004} value, 1.76.}
Similar to above, the $q = 0.2$ and $\eta_s = 0$ VS contact value predictions are in good agreement with the simulation results.
Generally, the $q = 0.1$ and 0.05 predictions are compatible with our data for $\eta_s \lesssim 0.1$ and 0.05, respectively, similar to the ranges shown in Fig.~\ref{fig:fig5}.

To showcase the utility of the computational framework used in this study, Fig.~\ref{fig:fig7} shows results of simulations sweeping over $q$ for $\eta_\ell = 0.35$ and fixed $\eta_s = 0.02$ [panel (a)] and $\eta_s = 0.06$ [panel (b)].
Results for $q = 0.05-0.2$ are reproduced from Fig.~\ref{fig:fig5}.
To our knowledge, the range of $q$ values considered surpasses those of any explicit binary mixture simulation technique to date.
The largest of these systems, $q = 0.02$ ($\sigma_\ell = 50$) and $\eta_s = 0.02$, contains $3.6\times 10^6$ particles; each subsequent halving of $q$ requires an $8\times$ increase in the number of small particles to maintain a constant value of $\eta_s$.
Aside from the behavior at contact, Fig.~\ref{fig:fig7}(a) shows no noticeable changes in the RDFs with decreasing $q$, implying that these mixtures act like single component fluids outside of contact.
The most prominent change is an essentially exponential increase in $g_\ell(\sigma_\ell)$ with increasing $\sigma_\ell$ (Fig.~\ref{fig:fig7} inset).
In the colloidal limit, any finite volume fraction of small particles produces the fluid-solid coexistence phase~\cite{dijkstra1999}, thus for sufficiently small $q$ more pronounced correlations should emerge for $\eta_s = 0.02$.
The $\eta_s = 0.06$ case, shown in Fig.~\ref{fig:fig7}(b), differs from that of $\eta_s = 0.02$ in that all RDFs are clearly distinct over the radial range depicted, and thus the mixtures do not act like single component fluids.
Indeed, more structure is evident near contact and at $r = 2\sigma_\ell$.
A single local minimum and maximum between $r = \sigma_\ell$ and $r = 1.1\sigma_\ell$ forms, implying that the depletion potential adopts a small repulsive barrier near contact.

For $\eta_s = 0.02$, the VS contact value predictions are surprisingly close to those estimated from the simulation data over the full range, despite the particle size disparity and the proximity of the fluid-solid phase boundary for $q = 0.0\bar{3}$ and below, which Ref.~\cite{dijkstra1999} predicts to be located at $\eta_s \leq 0.037$.
However, for $\eta_s = 0.06$ only the $q = 0.2$ and 0.1 values agree, and the VS expression under predicts $g_\ell(\sigma_\ell)$ for systems approaching or nominally within the fluid-solid coexistence region, similar to the behavior illustrated in Fig.~\ref{fig:fig5} and Fig.~\ref{fig:fig6}.

Explicit simulation of all particles permits the calculation of the inter-species RDF $g_{\ell s}(r)$ in addition to $g_\ell(r)$.
Figure~\ref{fig:fig8} shows $g_{\ell s}(r)$ for $q = 0.2$ and 0.1, with the particle separation shifted by the average particle diameter of the two species $\sigma_{\ell s}$.
The main features exhibited by these RDFs bear strong resemblance to those of the large-large RDFs, i.e., increasing contact values, oscillations in $g_{\ell s}(r)$ just outside of contact, and slight sharpening of a secondary peak, in this case located at $r = \sigma_{\ell s}+\sigma_\ell$.
The oscillations are shown most clearly for $\eta_s = 0.16$ and are spaced by $\sigma_s$ as above.
The contact value increases more gradually with $\eta_s$ than was observed for $g_\ell(\sigma_\ell)$, and the agreement with the VS predictions is somewhat poorer overall compared to the corresponding $g_\ell(\sigma_\ell)$ predictions.

\subsection{Large particle structure factors}
\label{sec:sk}
The static structure factor $S(k)$ is computed by summing over the $N_\ell$ large particle positions as $S(k) =  \sum_{m,n}^{N_\ell}\langle\text{exp}\left[i\mathbf{k}\cdot(\mathbf{r}_m-\mathbf{r}_n)\right]\rangle/N_\ell$.
Due to the periodic boundary conditions, the wavevectors ${\bf k}$ are limited to ${\bf k} = 2\pi/L (n_x,n_y,n_z)$, where $L$ is the length of the simulation cell and
$n_x$, $n_y$, and $n_z$ are integers.
Results for $S(k = |\mathbf{k}|)$ are shown in Fig.~\ref{fig:fig9} and Fig.~\ref{fig:fig10} for $\eta_\ell = 0.35$ and $0.20$.
The $S(k)$ results strongly mirror those of~\citet{dijkstra1999}, which were obtained using an effective depletion interaction between large particles (c.f., Fig. 10 and Fig. 12 of Ref.~\cite{dijkstra1999}).
In each case, data for the corresponding single component fluid are plotted for comparison.
The general trend of $S(k)$ with increasing $\eta_s$ evident in both figures is a successive  shift to higher $k$ of the peaks located near even integer multiples of $\pi$ and of the troughs located at odd integer multiples.
This shift was attributed to interaction potentials with short-ranged attraction~\cite{dijkstra1999}, and indeed is most pronounced for systems where $g_\ell(r)$ possesses a local minimum just outside of contact.
Initially, the primary peak near $2\pi$ reduces in height compared to the single component fluid, but at higher fractions of small particles, i.e, those corresponding to systems well within the fluid-solid coexistence region, the peak increases again, as do the second and third peaks.
Similar to the $S(k)$ shown in Ref.~\cite{dijkstra1999}, $\eta_\ell$ dictates the height of the primary and subsequent peaks: decreasing from $\eta_\ell = 0.35$ to 0.20 reduces the magnitude of the peaks.
At higher $k\sigma_\ell$ (not shown), $S(k)\rightarrow 1$ and higher order peaks continually decrease in magnitude.

The other main feature of $S(k)$, also discussed by~\citet{dijkstra1999}, is the behavior of $S(k\rightarrow 0)$ with increasing $\eta_s$.
In all cases shown in Fig.~\ref{fig:fig9} and Fig.~\ref{fig:fig10}, $S(k)$ for the smallest $k$ increases, signifying that the strength of the depletion forces grows with increasing $\eta_s$~\cite{gotzelmann1998,dijkstra1999}.
The overall magnitude of the lowest $S(k)$ is higher for $\eta_\ell = 0.20$ than for $\eta_\ell = 0.35$, consistent with the notion that oscillations about unity are more muted for the former.
For $\eta_\ell = 0.35$, $S(k\rightarrow 0)$ for the single highest $\eta_s$ values for $q = 0.1$ and $q = 0.05$ show non-monotonic behavior with decreasing $k$.
Note that the data are generally more noisy in this regime as a result of the scarcity of $k$ corresponding to wavelengths of order $L$, but in these two specific cases the non-monotonicity is clear.
An upturn at low $k$ is a signature of clusters of large particles forming, and is consistent with these two systems residing deep within the fluid-solid coexistence region.
However, as described above, large particle clusters formed during the simulations are transient.

\begin{figure}
\centering
\includegraphics[width=0.4\textwidth]{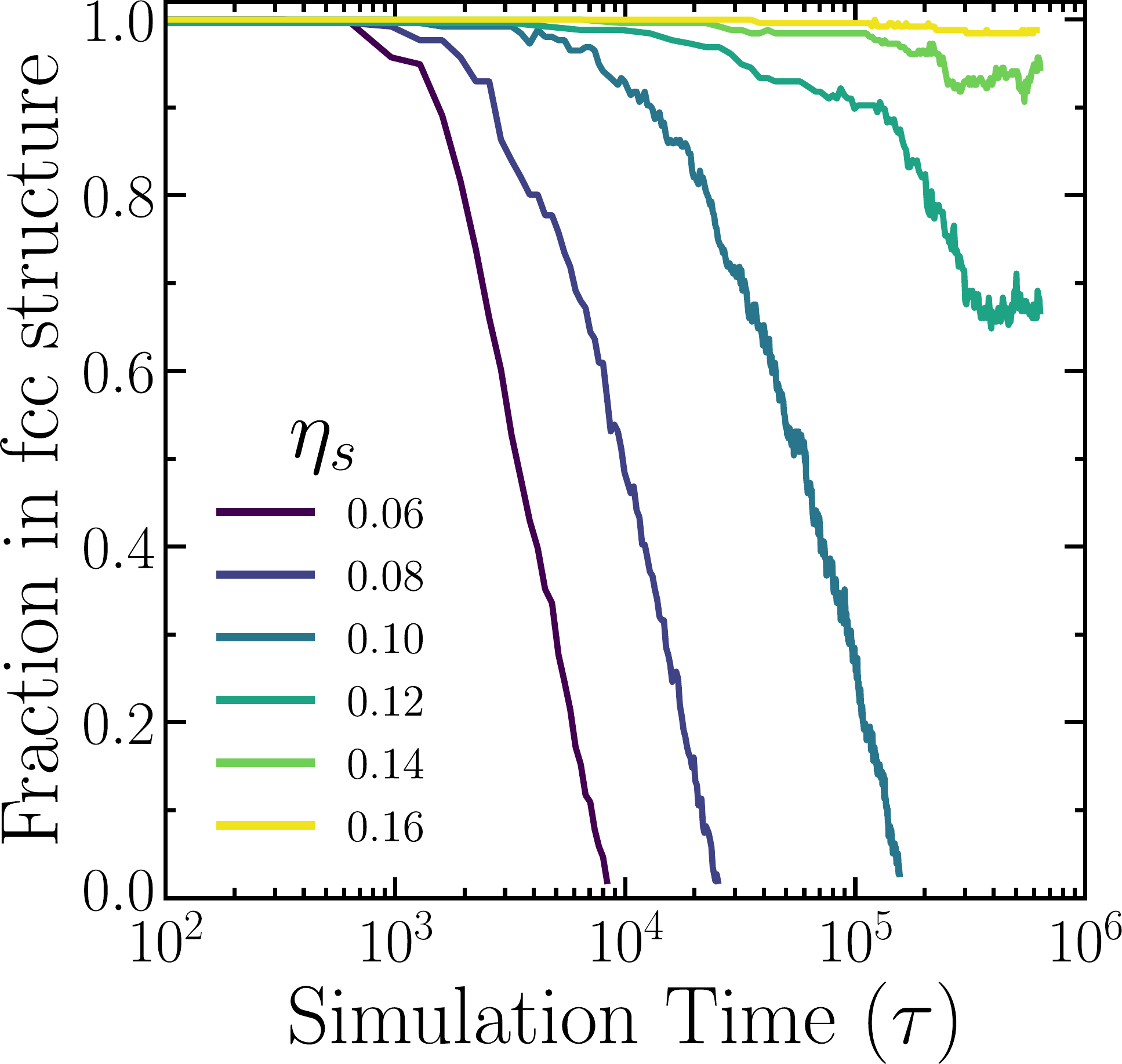}
\caption{{Number fraction of large particles identified as having fcc structure for $q = 0.1$ and $\eta_{\ell} = 0.35$ for the indicated $\eta_{s}$.}}
\label{fig:fig11}
\end{figure}
{
\subsection{Stability of fcc crystallites}
\label{sec:stability}
The absence of large particle crystal formation for state points nominally in the fluid-solid coexistence region may be a result of system size and ergodicity limitations in our simulations.
To overcome these limitations, large particles can be arranged directly into an fcc crystallite at the outset (instead of random placement) and the lifetime of the crystallite subsequently tracked as the simulation progresses.
We performed lengthy simulations to evaluate the stability of $5\times5\times5$ unit cell large particle crystallites with $q = 0.1$ and $\eta_\ell = 0.35$ for a range of $\eta_s$ that traversed the corresponding phase boundary, predicted to be at $\eta_s \approx 0.07$.
Figure~\ref{fig:fig11} shows the number fraction of (non-surface) large particles identified as belonging to the fcc structure by the Ackland-Jones technique~\cite{ackland2006} over time.
For $\eta_s \leq 0.10$, the crystallite fully melted over increasingly long time spans, implying that $\eta_s$ is too small for the fluid and solid phases to coexist.
However, for $\eta_s \geq 0.12$, at least a portion of the crystallite persisted to long simulation times and the number fraction appears to plateau, signaling coexistence between the two phases.
Steady state variations in the fcc fraction occur due to equilibrium between particles being dislodged from the crystallite and reattaching some time later.
For $\eta_s = 0.16$, only particles placed near the cube corners, which have the fewest neighbors, are dislodged, and the crystallite rotates as a rigid object in the small particle fluid.
Note that for  $\eta_s = 0.16$, with total particle density $\eta = 0.51$, the dynamics of both small and large particles are quite slow, so the fcc fraction shown in Fig.~\ref{fig:fig11} may not represent the equilibrium number fraction of fluid-solid coexistence.

The discrepancy between the predicted location of the phase boundary at $\eta_s \approx 0.07$ and our results indicating $\eta_s \approx 0.10-0.12$ may arise from the small number of large particles employed in the simulations.
Predicted coexistence densities from ~\citet{dijkstra1999} (see Fig. 15 of that work) imply that most large particles are in the fluid phase for $\eta_\ell = 0.35$ and $0.07 \leq \eta_s \leq 0.10$ (i.e., the large particle fluid coexistence density is close to $\eta_\ell$), with only a small number fraction of large particles expected to exist at the fcc density $\sim 0.74$.
This suggests that the equilibrium crystallite size for systems at or just above the phase boundary likely contains too few unit cells for $N_\ell = 500$ to be stable with respect to fluctuations.
We hypothesize that for $\eta_s \gtrsim 0.10$, the large particle fluid coexistence density is low enough compared to $\eta_\ell$ that the number fraction of particles in the crystallite is appreciable and measurable via simulation, leading to the overestimate of $\eta_s$ for the phase boundary that we observe.
Thus, simulations employing significantly greater large and overall particle counts are needed to confirm numerical predictions for the phase boundary.
}

\section{Conclusions}
Large-scale molecular dynamics simulations were performed leveraging a recently implemented neighbor-binning algorithm that permits the explicit numerical treatment of highly asymmetric binary fluid mixtures of spherical particles.
A stiff, linearly repulsive spring interaction acting between particles was employed to approach the hard sphere limit.
The simulations were conducted over diffusive timescales and were used to extract large-large and large-small radial distribution functions, their associated contact values, and large-large structure factors.
The results compared favorably with previous numerical work involving binary hard sphere mixtures with relatively modest particle size disparity and with other simulation approaches employing effective depletion interactions.
Comparisons with predictions from empirical expressions for the large-large and large-small radial distribution function contact values were also shown.
The simulations probed overall particle volume fractions that were largely unexplored, including volume fractions located within the predicted fluid-solid coexistence region, and considered particle size disparities greater than can be found in previous simulation studies of binary fluid mixtures to date.
{The stability of the predicted fcc crystal near the fluid-solid phase boundary was also tested}.
The simulation capability described in this work opens the door to more robust numerical treatment of particle interactions in the colloidal limit, including inter-particle friction and cohesion/adhesion, and arbitrary particle size distributions, without necessitating the use of effective interactions obtained by integrating out small particle degrees of freedom.

\section{Acknowledgements}
This work was performed at the Center for Integrated Nanotechnologies, a U.S. DOE and Office of Basic Energy Sciences user facility.
Sandia National Laboratories is a multi-mission laboratory managed and operated by National Technology and Engineering Solutions of Sandia, LLC., a wholly owned subsidiary of Honeywell International, Inc., for the U.S. Department of Energy National Nuclear Security Administration under contract DE-NA-0003525.
The views expressed in the article do not necessarily represent the views of the U.S. Department of Energy or the United States Government.

\bibliography{bib}

\begin{thebibliography}{54}%
\makeatletter
\providecommand \@ifxundefined [1]{%
 \@ifx{#1\undefined}
}%
\providecommand \@ifnum [1]{%
 \ifnum #1\expandafter \@firstoftwo
 \else \expandafter \@secondoftwo
 \fi
}%
\providecommand \@ifx [1]{%
 \ifx #1\expandafter \@firstoftwo
 \else \expandafter \@secondoftwo
 \fi
}%
\providecommand \natexlab [1]{#1}%
\providecommand \enquote  [1]{``#1''}%
\providecommand \bibnamefont  [1]{#1}%
\providecommand \bibfnamefont [1]{#1}%
\providecommand \citenamefont [1]{#1}%
\providecommand \href@noop [0]{\@secondoftwo}%
\providecommand \href [0]{\begingroup \@sanitize@url \@href}%
\providecommand \@href[1]{\@@startlink{#1}\@@href}%
\providecommand \@@href[1]{\endgroup#1\@@endlink}%
\providecommand \@sanitize@url [0]{\catcode `\\12\catcode `\$12\catcode
  `\&12\catcode `\#12\catcode `\^12\catcode `\_12\catcode `\%12\relax}%
\providecommand \@@startlink[1]{}%
\providecommand \@@endlink[0]{}%
\providecommand \url  [0]{\begingroup\@sanitize@url \@url }%
\providecommand \@url [1]{\endgroup\@href {#1}{\urlprefix }}%
\providecommand \urlprefix  [0]{URL }%
\providecommand \Eprint [0]{\href }%
\providecommand \doibase [0]{https://doi.org/}%
\providecommand \selectlanguage [0]{\@gobble}%
\providecommand \bibinfo  [0]{\@secondoftwo}%
\providecommand \bibfield  [0]{\@secondoftwo}%
\providecommand \translation [1]{[#1]}%
\providecommand \BibitemOpen [0]{}%
\providecommand \bibitemStop [0]{}%
\providecommand \bibitemNoStop [0]{.\EOS\space}%
\providecommand \EOS [0]{\spacefactor3000\relax}%
\providecommand \BibitemShut  [1]{\csname bibitem#1\endcsname}%
\let\auto@bib@innerbib\@empty
\bibitem [{\citenamefont {Lekkerkerker}\ \emph {et~al.}(1992)\citenamefont
  {Lekkerkerker}, \citenamefont {Poon}, \citenamefont {Pusey}, \citenamefont
  {Stroobants},\ and\ \citenamefont {Warren}}]{lekkerkerker1992}%
  \BibitemOpen
  \bibfield  {author} {\bibinfo {author} {\bibfnamefont {H.~N.~W.}\
  \bibnamefont {Lekkerkerker}}, \bibinfo {author} {\bibfnamefont {W.~C.-K.}\
  \bibnamefont {Poon}}, \bibinfo {author} {\bibfnamefont {P.~N.}\ \bibnamefont
  {Pusey}}, \bibinfo {author} {\bibfnamefont {A.}~\bibnamefont {Stroobants}},\
  and\ \bibinfo {author} {\bibfnamefont {P.~B.}\ \bibnamefont {Warren}},\
  }\href {https://doi.org/10.1209/0295-5075/20/6/015} {\bibfield  {journal}
  {\bibinfo  {journal} {Europhys. Lett.}\ }\textbf {\bibinfo {volume} {20}},\
  \bibinfo {pages} {559} (\bibinfo {year} {1992})}\BibitemShut {NoStop}%
\bibitem [{\citenamefont {Imhof}\ and\ \citenamefont
  {Dhont}(1995)}]{imhof1995}%
  \BibitemOpen
  \bibfield  {author} {\bibinfo {author} {\bibfnamefont {A.}~\bibnamefont
  {Imhof}}\ and\ \bibinfo {author} {\bibfnamefont {J.~K.~G.}\ \bibnamefont
  {Dhont}},\ }\href {https://doi.org/10.1103/PhysRevE.52.6344} {\bibfield
  {journal} {\bibinfo  {journal} {Phys. Rev. E}\ }\textbf {\bibinfo {volume}
  {52}},\ \bibinfo {pages} {6344} (\bibinfo {year} {1995})}\BibitemShut
  {NoStop}%
\bibitem [{\citenamefont {Zhu}\ \emph {et~al.}(1997)\citenamefont {Zhu},
  \citenamefont {Li}, \citenamefont {Rogers}, \citenamefont {Meyer},
  \citenamefont {Ottewill}, \citenamefont {Russel},\ and\ \citenamefont
  {Chaikin}}]{zhu1997}%
  \BibitemOpen
  \bibfield  {author} {\bibinfo {author} {\bibfnamefont {J.}~\bibnamefont
  {Zhu}}, \bibinfo {author} {\bibfnamefont {M.}~\bibnamefont {Li}}, \bibinfo
  {author} {\bibfnamefont {R.}~\bibnamefont {Rogers}}, \bibinfo {author}
  {\bibfnamefont {W.}~\bibnamefont {Meyer}}, \bibinfo {author} {\bibfnamefont
  {R.~H.}\ \bibnamefont {Ottewill}}, \bibinfo {author} {\bibfnamefont {W.~B.}\
  \bibnamefont {Russel}},\ and\ \bibinfo {author} {\bibfnamefont {P.~M.}\
  \bibnamefont {Chaikin}},\ }\href {https://doi.org/10.1038/43141} {\bibfield
  {journal} {\bibinfo  {journal} {Nature}\ }\textbf {\bibinfo {volume} {387}},\
  \bibinfo {pages} {883} (\bibinfo {year} {1997})}\BibitemShut {NoStop}%
\bibitem [{\citenamefont {Weeks}\ \emph {et~al.}(2000)\citenamefont {Weeks},
  \citenamefont {Crocker}, \citenamefont {Levitt}, \citenamefont {Schofield},\
  and\ \citenamefont {Weitz}}]{weeks2000}%
  \BibitemOpen
  \bibfield  {author} {\bibinfo {author} {\bibfnamefont {E.~R.}\ \bibnamefont
  {Weeks}}, \bibinfo {author} {\bibfnamefont {J.~C.}\ \bibnamefont {Crocker}},
  \bibinfo {author} {\bibfnamefont {A.~C.}\ \bibnamefont {Levitt}}, \bibinfo
  {author} {\bibfnamefont {A.}~\bibnamefont {Schofield}},\ and\ \bibinfo
  {author} {\bibfnamefont {D.~A.}\ \bibnamefont {Weitz}},\ }\href
  {https://doi.org/10.1126/science.287.5453.627} {\bibfield  {journal}
  {\bibinfo  {journal} {Science}\ }\textbf {\bibinfo {volume} {287}},\ \bibinfo
  {pages} {627} (\bibinfo {year} {2000})}\BibitemShut {NoStop}%
\bibitem [{\citenamefont {Royall}\ \emph {et~al.}(2013)\citenamefont {Royall},
  \citenamefont {Poon},\ and\ \citenamefont {Weeks}}]{royall2013}%
  \BibitemOpen
  \bibfield  {author} {\bibinfo {author} {\bibfnamefont {C.~P.}\ \bibnamefont
  {Royall}}, \bibinfo {author} {\bibfnamefont {W.~C.~K.}\ \bibnamefont
  {Poon}},\ and\ \bibinfo {author} {\bibfnamefont {E.~R.}\ \bibnamefont
  {Weeks}},\ }\href {https://doi.org/10.1039/C2SM26245B} {\bibfield  {journal}
  {\bibinfo  {journal} {Soft Matter}\ }\textbf {\bibinfo {volume} {9}},\
  \bibinfo {pages} {17} (\bibinfo {year} {2013})}\BibitemShut {NoStop}%
\bibitem [{\citenamefont {Furnas}(1931)}]{furnas1931}%
  \BibitemOpen
  \bibfield  {author} {\bibinfo {author} {\bibfnamefont {C.~C.}\ \bibnamefont
  {Furnas}},\ }\href {https://doi.org/10.1021/ie50261a017} {\bibfield
  {journal} {\bibinfo  {journal} {Ind. Eng. Chem. Res.}\ }\textbf {\bibinfo
  {volume} {23}},\ \bibinfo {pages} {1052} (\bibinfo {year}
  {1931})}\BibitemShut {NoStop}%
\bibitem [{\citenamefont {Prasad}\ \emph {et~al.}(2017)\citenamefont {Prasad},
  \citenamefont {Santangelo},\ and\ \citenamefont {Grason}}]{prasad2017}%
  \BibitemOpen
  \bibfield  {author} {\bibinfo {author} {\bibfnamefont {I.}~\bibnamefont
  {Prasad}}, \bibinfo {author} {\bibfnamefont {C.}~\bibnamefont {Santangelo}},\
  and\ \bibinfo {author} {\bibfnamefont {G.}~\bibnamefont {Grason}},\ }\href
  {https://doi.org/10.1103/PhysRevE.96.052905} {\bibfield  {journal} {\bibinfo
  {journal} {Phys. Rev. E}\ }\textbf {\bibinfo {volume} {96}},\ \bibinfo
  {pages} {052905} (\bibinfo {year} {2017})}\BibitemShut {NoStop}%
\bibitem [{\citenamefont {Srivastava}\ \emph {et~al.}(2021)\citenamefont
  {Srivastava}, \citenamefont {Roberts}, \citenamefont {Clemmer}, \citenamefont
  {Silbert}, \citenamefont {Lechman},\ and\ \citenamefont
  {Grest}}]{srivastava2021}%
  \BibitemOpen
  \bibfield  {author} {\bibinfo {author} {\bibfnamefont {I.}~\bibnamefont
  {Srivastava}}, \bibinfo {author} {\bibfnamefont {S.~A.}\ \bibnamefont
  {Roberts}}, \bibinfo {author} {\bibfnamefont {J.~T.}\ \bibnamefont
  {Clemmer}}, \bibinfo {author} {\bibfnamefont {L.~E.}\ \bibnamefont
  {Silbert}}, \bibinfo {author} {\bibfnamefont {J.~B.}\ \bibnamefont
  {Lechman}},\ and\ \bibinfo {author} {\bibfnamefont {G.~S.}\ \bibnamefont
  {Grest}},\ }\href {https://doi.org/10.1103/PhysRevResearch.3.L032042}
  {\bibfield  {journal} {\bibinfo  {journal} {Phys. Rev. Res.}\ }\textbf
  {\bibinfo {volume} {3}},\ \bibinfo {pages} {L032042} (\bibinfo {year}
  {2021})}\BibitemShut {NoStop}%
\bibitem [{\citenamefont {Pusey}\ and\ \citenamefont {van
  Megen}(1986)}]{pusey1986}%
  \BibitemOpen
  \bibfield  {author} {\bibinfo {author} {\bibfnamefont {P.~N.}\ \bibnamefont
  {Pusey}}\ and\ \bibinfo {author} {\bibfnamefont {W.}~\bibnamefont {van
  Megen}},\ }\href {https://doi.org/10.1038/320340a0} {\bibfield  {journal}
  {\bibinfo  {journal} {Nature}\ }\textbf {\bibinfo {volume} {320}},\ \bibinfo
  {pages} {340} (\bibinfo {year} {1986})}\BibitemShut {NoStop}%
\bibitem [{\citenamefont {Biben}\ and\ \citenamefont
  {Hansen}(1991)}]{biben1991}%
  \BibitemOpen
  \bibfield  {author} {\bibinfo {author} {\bibfnamefont {T.}~\bibnamefont
  {Biben}}\ and\ \bibinfo {author} {\bibfnamefont {J.-P.}\ \bibnamefont
  {Hansen}},\ }\href {https://doi.org/10.1103/PhysRevLett.66.2215} {\bibfield
  {journal} {\bibinfo  {journal} {Phys. Rev. Lett.}\ }\textbf {\bibinfo
  {volume} {66}},\ \bibinfo {pages} {2215} (\bibinfo {year}
  {1991})}\BibitemShut {NoStop}%
\bibitem [{\citenamefont {Rosenfeld}(1994)}]{rosenfeld1994}%
  \BibitemOpen
  \bibfield  {author} {\bibinfo {author} {\bibfnamefont {Y.}~\bibnamefont
  {Rosenfeld}},\ }\href {https://doi.org/10.1103/PhysRevLett.72.3831}
  {\bibfield  {journal} {\bibinfo  {journal} {Phys. Rev. Lett.}\ }\textbf
  {\bibinfo {volume} {72}},\ \bibinfo {pages} {3831} (\bibinfo {year}
  {1994})}\BibitemShut {NoStop}%
\bibitem [{\citenamefont {Dijkstra}\ \emph {et~al.}(1999)\citenamefont
  {Dijkstra}, \citenamefont {van Roij},\ and\ \citenamefont
  {Evans}}]{dijkstra1999}%
  \BibitemOpen
  \bibfield  {author} {\bibinfo {author} {\bibfnamefont {M.}~\bibnamefont
  {Dijkstra}}, \bibinfo {author} {\bibfnamefont {R.}~\bibnamefont {van Roij}},\
  and\ \bibinfo {author} {\bibfnamefont {R.}~\bibnamefont {Evans}},\ }\href
  {https://doi.org/10.1103/PhysRevE.59.5744} {\bibfield  {journal} {\bibinfo
  {journal} {Phys. Rev. E}\ }\textbf {\bibinfo {volume} {59}},\ \bibinfo
  {pages} {5744} (\bibinfo {year} {1999})}\BibitemShut {NoStop}%
\bibitem [{\citenamefont {Attard}(1989)}]{attard1989}%
  \BibitemOpen
  \bibfield  {author} {\bibinfo {author} {\bibfnamefont {P.}~\bibnamefont
  {Attard}},\ }\href {https://doi.org/10.1063/1.456931} {\bibfield  {journal}
  {\bibinfo  {journal} {J. Chem. Phys.}\ }\textbf {\bibinfo {volume} {91}},\
  \bibinfo {pages} {3083} (\bibinfo {year} {1989})}\BibitemShut {NoStop}%
\bibitem [{\citenamefont {Biben}\ \emph {et~al.}(1996)\citenamefont {Biben},
  \citenamefont {Bladon},\ and\ \citenamefont {Frenkel}}]{biben1996}%
  \BibitemOpen
  \bibfield  {author} {\bibinfo {author} {\bibfnamefont {T.}~\bibnamefont
  {Biben}}, \bibinfo {author} {\bibfnamefont {P.}~\bibnamefont {Bladon}},\ and\
  \bibinfo {author} {\bibfnamefont {D.}~\bibnamefont {Frenkel}},\ }\href
  {https://doi.org/10.1088/0953-8984/8/50/008} {\bibfield  {journal} {\bibinfo
  {journal} {J. Phys.: Condens. Matter}\ }\textbf {\bibinfo {volume} {8}},\
  \bibinfo {pages} {10799} (\bibinfo {year} {1996})}\BibitemShut {NoStop}%
\bibitem [{\citenamefont {Dickman}\ \emph {et~al.}(1997)\citenamefont
  {Dickman}, \citenamefont {Attard},\ and\ \citenamefont
  {Simonian}}]{dickman1997}%
  \BibitemOpen
  \bibfield  {author} {\bibinfo {author} {\bibfnamefont {R.}~\bibnamefont
  {Dickman}}, \bibinfo {author} {\bibfnamefont {P.}~\bibnamefont {Attard}},\
  and\ \bibinfo {author} {\bibfnamefont {V.}~\bibnamefont {Simonian}},\ }\href
  {https://doi.org/10.1063/1.474367} {\bibfield  {journal} {\bibinfo  {journal}
  {J. Chem. Phys.}\ }\textbf {\bibinfo {volume} {107}},\ \bibinfo {pages} {205}
  (\bibinfo {year} {1997})}\BibitemShut {NoStop}%
\bibitem [{\citenamefont {G{\"{o}}tzelmann}\ \emph {et~al.}(1998)\citenamefont
  {G{\"{o}}tzelmann}, \citenamefont {Evans},\ and\ \citenamefont
  {Dietrich}}]{gotzelmann1998}%
  \BibitemOpen
  \bibfield  {author} {\bibinfo {author} {\bibfnamefont {B.}~\bibnamefont
  {G{\"{o}}tzelmann}}, \bibinfo {author} {\bibfnamefont {R.}~\bibnamefont
  {Evans}},\ and\ \bibinfo {author} {\bibfnamefont {S.}~\bibnamefont
  {Dietrich}},\ }\href {https://doi.org/10.1103/PhysRevE.57.6785} {\bibfield
  {journal} {\bibinfo  {journal} {Phys. Rev. E}\ }\textbf {\bibinfo {volume}
  {57}},\ \bibinfo {pages} {6785} (\bibinfo {year} {1998})}\BibitemShut
  {NoStop}%
\bibitem [{\citenamefont {Ashton}\ \emph {et~al.}(2011)\citenamefont {Ashton},
  \citenamefont {Wilding}, \citenamefont {Roth},\ and\ \citenamefont
  {Evans}}]{ashton2011}%
  \BibitemOpen
  \bibfield  {author} {\bibinfo {author} {\bibfnamefont {D.~J.}\ \bibnamefont
  {Ashton}}, \bibinfo {author} {\bibfnamefont {N.~B.}\ \bibnamefont {Wilding}},
  \bibinfo {author} {\bibfnamefont {R.}~\bibnamefont {Roth}},\ and\ \bibinfo
  {author} {\bibfnamefont {R.}~\bibnamefont {Evans}},\ }\href
  {https://doi.org/10.1103/PhysRevE.84.061136} {\bibfield  {journal} {\bibinfo
  {journal} {Phys. Rev. E}\ }\textbf {\bibinfo {volume} {84}},\ \bibinfo
  {pages} {061136} (\bibinfo {year} {2011})}\BibitemShut {NoStop}%
\bibitem [{\citenamefont {Lue}\ and\ \citenamefont {Woodcock}(1999)}]{lue1999}%
  \BibitemOpen
  \bibfield  {author} {\bibinfo {author} {\bibfnamefont {L.}~\bibnamefont
  {Lue}}\ and\ \bibinfo {author} {\bibfnamefont {L.~V.}\ \bibnamefont
  {Woodcock}},\ }\href {https://doi.org/10.1080/00268979909483087} {\bibfield
  {journal} {\bibinfo  {journal} {Mol. Phys.}\ }\textbf {\bibinfo {volume}
  {96}},\ \bibinfo {pages} {1435} (\bibinfo {year} {1999})}\BibitemShut
  {NoStop}%
\bibitem [{\citenamefont {Malherbe}\ and\ \citenamefont
  {Amokrane}(2001)}]{malherbe2001}%
  \BibitemOpen
  \bibfield  {author} {\bibinfo {author} {\bibfnamefont {J.~G.}\ \bibnamefont
  {Malherbe}}\ and\ \bibinfo {author} {\bibfnamefont {S.}~\bibnamefont
  {Amokrane}},\ }\href {https://doi.org/10.1080/00268970010012617} {\bibfield
  {journal} {\bibinfo  {journal} {Mol. Phys.}\ }\textbf {\bibinfo {volume}
  {99}},\ \bibinfo {pages} {355} (\bibinfo {year} {2001})}\BibitemShut
  {NoStop}%
\bibitem [{\citenamefont {Grest}\ \emph {et~al.}(2011)\citenamefont {Grest},
  \citenamefont {Wang}, \citenamefont {in't Veld},\ and\ \citenamefont
  {Keffer}}]{grest2011}%
  \BibitemOpen
  \bibfield  {author} {\bibinfo {author} {\bibfnamefont {G.~S.}\ \bibnamefont
  {Grest}}, \bibinfo {author} {\bibfnamefont {Q.}~\bibnamefont {Wang}},
  \bibinfo {author} {\bibfnamefont {P.}~\bibnamefont {in't Veld}},\ and\
  \bibinfo {author} {\bibfnamefont {D.~J.}\ \bibnamefont {Keffer}},\ }\href
  {https://doi.org/10.1063/1.3578181} {\bibfield  {journal} {\bibinfo
  {journal} {J. Chem. Phys.}\ }\textbf {\bibinfo {volume} {134}},\ \bibinfo
  {pages} {144902} (\bibinfo {year} {2011})}\BibitemShut {NoStop}%
\bibitem [{\citenamefont {Kobayashi}\ \emph {et~al.}(2021)\citenamefont
  {Kobayashi}, \citenamefont {Rohrbach}, \citenamefont {Scheichl},
  \citenamefont {Wilding},\ and\ \citenamefont {Jack}}]{kobayashi2021}%
  \BibitemOpen
  \bibfield  {author} {\bibinfo {author} {\bibfnamefont {H.}~\bibnamefont
  {Kobayashi}}, \bibinfo {author} {\bibfnamefont {P.~B.}\ \bibnamefont
  {Rohrbach}}, \bibinfo {author} {\bibfnamefont {R.}~\bibnamefont {Scheichl}},
  \bibinfo {author} {\bibfnamefont {N.~B.}\ \bibnamefont {Wilding}},\ and\
  \bibinfo {author} {\bibfnamefont {R.~L.}\ \bibnamefont {Jack}},\ }\href
  {https://doi.org/10.1103/PhysRevE.104.044603} {\bibfield  {journal} {\bibinfo
   {journal} {Phys. Rev. E}\ }\textbf {\bibinfo {volume} {104}},\ \bibinfo
  {pages} {044603} (\bibinfo {year} {2021})}\BibitemShut {NoStop}%
\bibitem [{\citenamefont {Henderson}\ \emph {et~al.}(2005)\citenamefont
  {Henderson}, \citenamefont {Trokhymchuk}, \citenamefont {Woodcock},\ and\
  \citenamefont {Chan}}]{henderson2005}%
  \BibitemOpen
  \bibfield  {author} {\bibinfo {author} {\bibfnamefont {D.}~\bibnamefont
  {Henderson}}, \bibinfo {author} {\bibfnamefont {A.}~\bibnamefont
  {Trokhymchuk}}, \bibinfo {author} {\bibfnamefont {L.~V.}\ \bibnamefont
  {Woodcock}},\ and\ \bibinfo {author} {\bibfnamefont {K.-Y.}\ \bibnamefont
  {Chan}},\ }\href {https://doi.org/10.1080/00268970412531329082} {\bibfield
  {journal} {\bibinfo  {journal} {Mol. Phys.}\ }\textbf {\bibinfo {volume}
  {103}},\ \bibinfo {pages} {667} (\bibinfo {year} {2005})}\BibitemShut
  {NoStop}%
\bibitem [{\citenamefont {Alawneh}\ and\ \citenamefont
  {Henderson}(2008)}]{alawneh2008}%
  \BibitemOpen
  \bibfield  {author} {\bibinfo {author} {\bibfnamefont {M.}~\bibnamefont
  {Alawneh}}\ and\ \bibinfo {author} {\bibfnamefont {D.}~\bibnamefont
  {Henderson}},\ }\href {https://doi.org/10.1080/00268970802116906} {\bibfield
  {journal} {\bibinfo  {journal} {Mol. Phys.}\ }\textbf {\bibinfo {volume}
  {106}},\ \bibinfo {pages} {607} (\bibinfo {year} {2008})}\BibitemShut
  {NoStop}%
\bibitem [{\citenamefont {L{\'{a}}zaro-L{\'{a}}zaro}\ \emph
  {et~al.}(2019)\citenamefont {L{\'{a}}zaro-L{\'{a}}zaro}, \citenamefont
  {Perera-Burgos}, \citenamefont {Laermann}, \citenamefont {Sentjabrskaja},
  \citenamefont {P{\'{e}}rez-{\'{A}}ngel}, \citenamefont {Laurati},
  \citenamefont {Egelhaaf}, \citenamefont {Medina-Noyola}, \citenamefont
  {Voigtmann}, \citenamefont {Casta{\~{n}}eda-Priego},\ and\ \citenamefont
  {Elizondo-Aguilera}}]{lazaro2019}%
  \BibitemOpen
  \bibfield  {author} {\bibinfo {author} {\bibfnamefont {E.}~\bibnamefont
  {L{\'{a}}zaro-L{\'{a}}zaro}}, \bibinfo {author} {\bibfnamefont {J.~A.}\
  \bibnamefont {Perera-Burgos}}, \bibinfo {author} {\bibfnamefont
  {P.}~\bibnamefont {Laermann}}, \bibinfo {author} {\bibfnamefont
  {T.}~\bibnamefont {Sentjabrskaja}}, \bibinfo {author} {\bibfnamefont
  {G.}~\bibnamefont {P{\'{e}}rez-{\'{A}}ngel}}, \bibinfo {author}
  {\bibfnamefont {M.}~\bibnamefont {Laurati}}, \bibinfo {author} {\bibfnamefont
  {S.~U.}\ \bibnamefont {Egelhaaf}}, \bibinfo {author} {\bibfnamefont
  {M.}~\bibnamefont {Medina-Noyola}}, \bibinfo {author} {\bibfnamefont
  {T.}~\bibnamefont {Voigtmann}}, \bibinfo {author} {\bibfnamefont
  {R.}~\bibnamefont {Casta{\~{n}}eda-Priego}},\ and\ \bibinfo {author}
  {\bibfnamefont {L.~F.}\ \bibnamefont {Elizondo-Aguilera}},\ }\href
  {https://doi.org/10.1103/PhysRevE.99.042603} {\bibfield  {journal} {\bibinfo
  {journal} {Phys. Rev. E}\ }\textbf {\bibinfo {volume} {99}},\ \bibinfo
  {pages} {042603} (\bibinfo {year} {2019})}\BibitemShut {NoStop}%
\bibitem [{\citenamefont {Bommineni}\ \emph {et~al.}(2020)\citenamefont
  {Bommineni}, \citenamefont {Klement},\ and\ \citenamefont
  {Engel}}]{bommineni2020}%
  \BibitemOpen
  \bibfield  {author} {\bibinfo {author} {\bibfnamefont {P.~K.}\ \bibnamefont
  {Bommineni}}, \bibinfo {author} {\bibfnamefont {M.}~\bibnamefont {Klement}},\
  and\ \bibinfo {author} {\bibfnamefont {M.}~\bibnamefont {Engel}},\ }\href
  {https://doi.org/10.1103/PhysRevLett.124.218003} {\bibfield  {journal}
  {\bibinfo  {journal} {Phys. Rev. Lett.}\ }\textbf {\bibinfo {volume} {124}},\
  \bibinfo {pages} {218003} (\bibinfo {year} {2020})}\BibitemShut {NoStop}%
\bibitem [{\citenamefont {Pieprzyk}\ \emph {et~al.}(2020)\citenamefont
  {Pieprzyk}, \citenamefont {Bra{\'{n}}ka}, \citenamefont {Yuste},
  \citenamefont {Santos},\ and\ \citenamefont {de~Haro}}]{pieprzyk2020}%
  \BibitemOpen
  \bibfield  {author} {\bibinfo {author} {\bibfnamefont {S.}~\bibnamefont
  {Pieprzyk}}, \bibinfo {author} {\bibfnamefont {A.~C.}\ \bibnamefont
  {Bra{\'{n}}ka}}, \bibinfo {author} {\bibfnamefont {S.~B.}\ \bibnamefont
  {Yuste}}, \bibinfo {author} {\bibfnamefont {A.}~\bibnamefont {Santos}},\ and\
  \bibinfo {author} {\bibfnamefont {M.~L.}\ \bibnamefont {de~Haro}},\ }\href
  {https://doi.org/10.1103/PhysRevE.101.012117} {\bibfield  {journal} {\bibinfo
   {journal} {Phys. Rev. E}\ }\textbf {\bibinfo {volume} {101}},\ \bibinfo
  {pages} {012117} (\bibinfo {year} {2020})}\BibitemShut {NoStop}%
\bibitem [{\citenamefont {Pieprzyk}\ \emph {et~al.}(2021)\citenamefont
  {Pieprzyk}, \citenamefont {Yuste}, \citenamefont {Santos}, \citenamefont
  {de~Haro},\ and\ \citenamefont {Bra{\'{n}}ka}}]{pieprzyk2021}%
  \BibitemOpen
  \bibfield  {author} {\bibinfo {author} {\bibfnamefont {S.}~\bibnamefont
  {Pieprzyk}}, \bibinfo {author} {\bibfnamefont {S.~B.}\ \bibnamefont {Yuste}},
  \bibinfo {author} {\bibfnamefont {A.}~\bibnamefont {Santos}}, \bibinfo
  {author} {\bibfnamefont {M.~L.}\ \bibnamefont {de~Haro}},\ and\ \bibinfo
  {author} {\bibfnamefont {A.~C.}\ \bibnamefont {Bra{\'{n}}ka}},\ }\href
  {https://doi.org/10.1103/PhysRevE.104.054142} {\bibfield  {journal} {\bibinfo
   {journal} {Phys. Rev. E}\ }\textbf {\bibinfo {volume} {104}},\ \bibinfo
  {pages} {054142} (\bibinfo {year} {2021})}\BibitemShut {NoStop}%
\bibitem [{\citenamefont {Dress}\ and\ \citenamefont
  {Krauth}(1995)}]{dress1995}%
  \BibitemOpen
  \bibfield  {author} {\bibinfo {author} {\bibfnamefont {C.}~\bibnamefont
  {Dress}}\ and\ \bibinfo {author} {\bibfnamefont {W.}~\bibnamefont {Krauth}},\
  }\href {https://doi.org/10.1088/0305-4470/28/23/001} {\bibfield  {journal}
  {\bibinfo  {journal} {J. Phys. A: Math. Gen.}\ }\textbf {\bibinfo {volume}
  {28}},\ \bibinfo {pages} {L597} (\bibinfo {year} {1995})}\BibitemShut
  {NoStop}%
\bibitem [{\citenamefont {Buhot}\ and\ \citenamefont
  {Krauth}(1998)}]{buhot1998}%
  \BibitemOpen
  \bibfield  {author} {\bibinfo {author} {\bibfnamefont {A.}~\bibnamefont
  {Buhot}}\ and\ \bibinfo {author} {\bibfnamefont {W.}~\bibnamefont {Krauth}},\
  }\href {https://doi.org/10.1103/PhysRevLett.80.3787} {\bibfield  {journal}
  {\bibinfo  {journal} {Phys. Rev. Lett.}\ }\textbf {\bibinfo {volume} {80}},\
  \bibinfo {pages} {3787} (\bibinfo {year} {1998})}\BibitemShut {NoStop}%
\bibitem [{\citenamefont {Malherbe}\ and\ \citenamefont
  {Krauth}(2007)}]{malherbe2007}%
  \BibitemOpen
  \bibfield  {author} {\bibinfo {author} {\bibfnamefont {J.~G.}\ \bibnamefont
  {Malherbe}}\ and\ \bibinfo {author} {\bibfnamefont {W.}~\bibnamefont
  {Krauth}},\ }\href {https://doi.org/10.1080/00268970701678907} {\bibfield
  {journal} {\bibinfo  {journal} {Mol. Phys.}\ }\textbf {\bibinfo {volume}
  {105}},\ \bibinfo {pages} {2393} (\bibinfo {year} {2007})}\BibitemShut
  {NoStop}%
\bibitem [{\citenamefont {Ogarko}\ and\ \citenamefont
  {Luding}(2012)}]{ogarko2012}%
  \BibitemOpen
  \bibfield  {author} {\bibinfo {author} {\bibfnamefont {V.}~\bibnamefont
  {Ogarko}}\ and\ \bibinfo {author} {\bibfnamefont {S.}~\bibnamefont
  {Luding}},\ }\href
  {https://doi.org/https://doi.org/10.1016/j.cpc.2011.12.019} {\bibfield
  {journal} {\bibinfo  {journal} {Comput. Phys. Commun.}\ }\textbf {\bibinfo
  {volume} {183}},\ \bibinfo {pages} {931} (\bibinfo {year}
  {2012})}\BibitemShut {NoStop}%
\bibitem [{\citenamefont {Krijgsman}\ \emph {et~al.}(2014)\citenamefont
  {Krijgsman}, \citenamefont {Ogarko},\ and\ \citenamefont
  {Luding}}]{krijgsman2014}%
  \BibitemOpen
  \bibfield  {author} {\bibinfo {author} {\bibfnamefont {D.}~\bibnamefont
  {Krijgsman}}, \bibinfo {author} {\bibfnamefont {V.}~\bibnamefont {Ogarko}},\
  and\ \bibinfo {author} {\bibfnamefont {S.}~\bibnamefont {Luding}},\ }\href
  {https://doi.org/10.1007/s40571-014-0020-9} {\bibfield  {journal} {\bibinfo
  {journal} {Comp. Part. Mech.}\ }\textbf {\bibinfo {volume} {1}},\ \bibinfo
  {pages} {357} (\bibinfo {year} {2014})}\BibitemShut {NoStop}%
\bibitem [{\citenamefont {Stratford}\ \emph {et~al.}(2018)\citenamefont
  {Stratford}, \citenamefont {Shire},\ and\ \citenamefont
  {Hanley}}]{stratford2018}%
  \BibitemOpen
  \bibfield  {author} {\bibinfo {author} {\bibfnamefont {K.}~\bibnamefont
  {Stratford}}, \bibinfo {author} {\bibfnamefont {T.}~\bibnamefont {Shire}},\
  and\ \bibinfo {author} {\bibfnamefont {K.}~\bibnamefont {Hanley}},\
  }\href@noop {} {\emph {\bibinfo {title} {Implementation of multi-level
  contact detection in {LAMMPS}}}},\ \bibinfo {type} {Tech. Rep.}\ \bibinfo
  {number} {eCSE12-09}\ (\bibinfo  {institution} {University of Edinburgh
  (United Kingdom)},\ \bibinfo {year} {2018})\BibitemShut {NoStop}%
\bibitem [{\citenamefont {Shire}\ \emph {et~al.}(2021)\citenamefont {Shire},
  \citenamefont {Hanley},\ and\ \citenamefont {Stratford}}]{shire2021}%
  \BibitemOpen
  \bibfield  {author} {\bibinfo {author} {\bibfnamefont {T.}~\bibnamefont
  {Shire}}, \bibinfo {author} {\bibfnamefont {K.~J.}\ \bibnamefont {Hanley}},\
  and\ \bibinfo {author} {\bibfnamefont {K.}~\bibnamefont {Stratford}},\ }\href
  {https://doi.org/10.1007/s40571-020-00361-2} {\bibfield  {journal} {\bibinfo
  {journal} {Comp. Part. Mech.}\ }\textbf {\bibinfo {volume} {8}},\ \bibinfo
  {pages} {653} (\bibinfo {year} {2021})}\BibitemShut {NoStop}%
\bibitem [{\citenamefont {Thompson}\ \emph {et~al.}(2022)\citenamefont
  {Thompson}, \citenamefont {Aktulga}, \citenamefont {Berger}, \citenamefont
  {Bolintineanu}, \citenamefont {Brown}, \citenamefont {Crozier}, \citenamefont
  {{in 't Veld}}, \citenamefont {Kohlmeyer}, \citenamefont {Moore},
  \citenamefont {Nguyen}, \citenamefont {Shan}, \citenamefont {Stevens},
  \citenamefont {Tranchida}, \citenamefont {Trott},\ and\ \citenamefont
  {Plimpton}}]{thompson2022}%
  \BibitemOpen
  \bibfield  {author} {\bibinfo {author} {\bibfnamefont {A.~P.}\ \bibnamefont
  {Thompson}}, \bibinfo {author} {\bibfnamefont {H.~M.}\ \bibnamefont
  {Aktulga}}, \bibinfo {author} {\bibfnamefont {R.}~\bibnamefont {Berger}},
  \bibinfo {author} {\bibfnamefont {D.~S.}\ \bibnamefont {Bolintineanu}},
  \bibinfo {author} {\bibfnamefont {W.~M.}\ \bibnamefont {Brown}}, \bibinfo
  {author} {\bibfnamefont {P.~S.}\ \bibnamefont {Crozier}}, \bibinfo {author}
  {\bibfnamefont {P.~J.}\ \bibnamefont {{in 't Veld}}}, \bibinfo {author}
  {\bibfnamefont {A.}~\bibnamefont {Kohlmeyer}}, \bibinfo {author}
  {\bibfnamefont {S.~G.}\ \bibnamefont {Moore}}, \bibinfo {author}
  {\bibfnamefont {T.~D.}\ \bibnamefont {Nguyen}}, \bibinfo {author}
  {\bibfnamefont {R.}~\bibnamefont {Shan}}, \bibinfo {author} {\bibfnamefont
  {M.~J.}\ \bibnamefont {Stevens}}, \bibinfo {author} {\bibfnamefont
  {J.}~\bibnamefont {Tranchida}}, \bibinfo {author} {\bibfnamefont
  {C.}~\bibnamefont {Trott}},\ and\ \bibinfo {author} {\bibfnamefont {S.~J.}\
  \bibnamefont {Plimpton}},\ }\href {https://doi.org/10.1016/j.cpc.2021.108171}
  {\bibfield  {journal} {\bibinfo  {journal} {Comput. Phys. Commun.}\ }\textbf
  {\bibinfo {volume} {271}},\ \bibinfo {pages} {108171} (\bibinfo {year}
  {2022})}\BibitemShut {NoStop}%
\bibitem [{\citenamefont {Monti}\ \emph {et~al.}(2022)\citenamefont {Monti},
  \citenamefont {Clemmer}, \citenamefont {Srivastava}, \citenamefont {Silbert},
  \citenamefont {Grest},\ and\ \citenamefont {Lechman}}]{monti2022}%
  \BibitemOpen
  \bibfield  {author} {\bibinfo {author} {\bibfnamefont {J.~M.}\ \bibnamefont
  {Monti}}, \bibinfo {author} {\bibfnamefont {J.~T.}\ \bibnamefont {Clemmer}},
  \bibinfo {author} {\bibfnamefont {I.}~\bibnamefont {Srivastava}}, \bibinfo
  {author} {\bibfnamefont {L.~E.}\ \bibnamefont {Silbert}}, \bibinfo {author}
  {\bibfnamefont {G.~S.}\ \bibnamefont {Grest}},\ and\ \bibinfo {author}
  {\bibfnamefont {J.~B.}\ \bibnamefont {Lechman}},\ }\href@noop {} {\bibfield
  {journal} {\bibinfo  {journal} {Phys. Rev. E}\ }\textbf {\bibinfo {volume}
  {106}},\ \bibinfo {pages} {in press} (\bibinfo {year} {2022})}\BibitemShut
  {NoStop}%
\bibitem [{\citenamefont {Stukowski}(2009)}]{ovito}%
  \BibitemOpen
  \bibfield  {author} {\bibinfo {author} {\bibfnamefont {A.}~\bibnamefont
  {Stukowski}},\ }\href {https://doi.org/10.1088/0965-0393/18/1/015012}
  {\bibfield  {journal} {\bibinfo  {journal} {Modell. Simul. Mater. Sci. Eng.}\
  }\textbf {\bibinfo {volume} {18}},\ \bibinfo {pages} {015012} (\bibinfo
  {year} {2009})}\BibitemShut {NoStop}%
\bibitem [{\citenamefont {Hoover}\ and\ \citenamefont
  {Ree}(1968)}]{hoover1968}%
  \BibitemOpen
  \bibfield  {author} {\bibinfo {author} {\bibfnamefont {W.~G.}\ \bibnamefont
  {Hoover}}\ and\ \bibinfo {author} {\bibfnamefont {F.~H.}\ \bibnamefont
  {Ree}},\ }\href {https://doi.org/10.1063/1.1670641} {\bibfield  {journal}
  {\bibinfo  {journal} {J. Chem. Phys.}\ }\textbf {\bibinfo {volume} {49}},\
  \bibinfo {pages} {3609} (\bibinfo {year} {1968})}\BibitemShut {NoStop}%
\bibitem [{\citenamefont {Woodcock}(1997)}]{woodcock1997}%
  \BibitemOpen
  \bibfield  {author} {\bibinfo {author} {\bibfnamefont {L.}~\bibnamefont
  {Woodcock}},\ }\href {https://doi.org/10.1038/385141a0} {\bibfield  {journal}
  {\bibinfo  {journal} {Nature}\ }\textbf {\bibinfo {volume} {385}},\ \bibinfo
  {pages} {141} (\bibinfo {year} {1997})}\BibitemShut {NoStop}%
\bibitem [{\citenamefont {Bolhuis}\ \emph {et~al.}(1997)\citenamefont
  {Bolhuis}, \citenamefont {Frenkel}, \citenamefont {Mau},\ and\ \citenamefont
  {Huse}}]{bolhuis1997}%
  \BibitemOpen
  \bibfield  {author} {\bibinfo {author} {\bibfnamefont {P.~G.}\ \bibnamefont
  {Bolhuis}}, \bibinfo {author} {\bibfnamefont {D.}~\bibnamefont {Frenkel}},
  \bibinfo {author} {\bibfnamefont {S.-C.}\ \bibnamefont {Mau}},\ and\ \bibinfo
  {author} {\bibfnamefont {D.~A.}\ \bibnamefont {Huse}},\ }\href
  {https://doi.org/10.1038/40779} {\bibfield  {journal} {\bibinfo  {journal}
  {Nature}\ }\textbf {\bibinfo {volume} {388}},\ \bibinfo {pages} {235}
  (\bibinfo {year} {1997})}\BibitemShut {NoStop}%
\bibitem [{\citenamefont {Auer}\ and\ \citenamefont
  {Frenkel}(2001)}]{auer2001}%
  \BibitemOpen
  \bibfield  {author} {\bibinfo {author} {\bibfnamefont {S.}~\bibnamefont
  {Auer}}\ and\ \bibinfo {author} {\bibfnamefont {D.}~\bibnamefont {Frenkel}},\
  }\href {https://doi.org/10.1038/35059035} {\bibfield  {journal} {\bibinfo
  {journal} {Nature}\ }\textbf {\bibinfo {volume} {409}},\ \bibinfo {pages}
  {1020} (\bibinfo {year} {2001})}\BibitemShut {NoStop}%
\bibitem [{\citenamefont {Filion}\ \emph {et~al.}(2010)\citenamefont {Filion},
  \citenamefont {Hermes}, \citenamefont {Ni},\ and\ \citenamefont
  {Dijkstra}}]{filion2010}%
  \BibitemOpen
  \bibfield  {author} {\bibinfo {author} {\bibfnamefont {L.}~\bibnamefont
  {Filion}}, \bibinfo {author} {\bibfnamefont {M.}~\bibnamefont {Hermes}},
  \bibinfo {author} {\bibfnamefont {R.}~\bibnamefont {Ni}},\ and\ \bibinfo
  {author} {\bibfnamefont {M.}~\bibnamefont {Dijkstra}},\ }\href
  {https://doi.org/10.1063/1.3506838} {\bibfield  {journal} {\bibinfo
  {journal} {J. Chem. Phys.}\ }\textbf {\bibinfo {volume} {133}},\ \bibinfo
  {pages} {244115} (\bibinfo {year} {2010})}\BibitemShut {NoStop}%
\bibitem [{\citenamefont {Ackland}\ and\ \citenamefont
  {Jones}(2006)}]{ackland2006}%
  \BibitemOpen
  \bibfield  {author} {\bibinfo {author} {\bibfnamefont {G.~J.}\ \bibnamefont
  {Ackland}}\ and\ \bibinfo {author} {\bibfnamefont {A.~P.}\ \bibnamefont
  {Jones}},\ }\href {https://doi.org/10.1103/PhysRevB.73.054104} {\bibfield
  {journal} {\bibinfo  {journal} {Phys. Rev. B}\ }\textbf {\bibinfo {volume}
  {73}},\ \bibinfo {pages} {054104} (\bibinfo {year} {2006})}\BibitemShut
  {NoStop}%
\bibitem [{\citenamefont {Grest}\ and\ \citenamefont
  {Kremer}(1986)}]{grest1986}%
  \BibitemOpen
  \bibfield  {author} {\bibinfo {author} {\bibfnamefont {G.~S.}\ \bibnamefont
  {Grest}}\ and\ \bibinfo {author} {\bibfnamefont {K.}~\bibnamefont {Kremer}},\
  }\href {https://doi.org/10.1103/PhysRevA.33.3628} {\bibfield  {journal}
  {\bibinfo  {journal} {Phys. Rev. A}\ }\textbf {\bibinfo {volume} {33}},\
  \bibinfo {pages} {3628} (\bibinfo {year} {1986})}\BibitemShut {NoStop}%
\bibitem [{\citenamefont {Wang}\ and\ \citenamefont {Brady}(2015)}]{wang2015}%
  \BibitemOpen
  \bibfield  {author} {\bibinfo {author} {\bibfnamefont {M.}~\bibnamefont
  {Wang}}\ and\ \bibinfo {author} {\bibfnamefont {J.~F.}\ \bibnamefont
  {Brady}},\ }\href {https://doi.org/10.1063/1.4913518} {\bibfield  {journal}
  {\bibinfo  {journal} {J. Chem. Phys.}\ }\textbf {\bibinfo {volume} {142}},\
  \bibinfo {pages} {094901} (\bibinfo {year} {2015})},\ \Eprint
  {https://arxiv.org/abs/1412.8122} {1412.8122} \BibitemShut {NoStop}%
\bibitem [{\citenamefont {Viduna}\ and\ \citenamefont
  {Smith}(2002{\natexlab{a}})}]{viduna2002a}%
  \BibitemOpen
  \bibfield  {author} {\bibinfo {author} {\bibfnamefont {D.}~\bibnamefont
  {Viduna}}\ and\ \bibinfo {author} {\bibfnamefont {W.~R.}\ \bibnamefont
  {Smith}},\ }\href {https://doi.org/10.1063/1.1486446} {\bibfield  {journal}
  {\bibinfo  {journal} {J. Chem. Phys.}\ }\textbf {\bibinfo {volume} {117}},\
  \bibinfo {pages} {1214} (\bibinfo {year} {2002}{\natexlab{a}})}\BibitemShut
  {NoStop}%
\bibitem [{\citenamefont {Viduna}\ and\ \citenamefont
  {Smith}(2002{\natexlab{b}})}]{viduna2002b}%
  \BibitemOpen
  \bibfield  {author} {\bibinfo {author} {\bibfnamefont {D.}~\bibnamefont
  {Viduna}}\ and\ \bibinfo {author} {\bibfnamefont {W.~R.}\ \bibnamefont
  {Smith}},\ }\href {https://doi.org/10.1080/00268970210145311} {\bibfield
  {journal} {\bibinfo  {journal} {Mol. Phys.}\ }\textbf {\bibinfo {volume}
  {100}},\ \bibinfo {pages} {2903} (\bibinfo {year}
  {2002}{\natexlab{b}})}\BibitemShut {NoStop}%
\bibitem [{\citenamefont {Hermes}\ and\ \citenamefont
  {Dijkstra}(2010)}]{hermes2010}%
  \BibitemOpen
  \bibfield  {author} {\bibinfo {author} {\bibfnamefont {M.}~\bibnamefont
  {Hermes}}\ and\ \bibinfo {author} {\bibfnamefont {M.}~\bibnamefont
  {Dijkstra}},\ }\href {https://doi.org/10.1209/0295-5075/89/38005} {\bibfield
  {journal} {\bibinfo  {journal} {Europhys. Lett.}\ }\textbf {\bibinfo {volume}
  {89}},\ \bibinfo {pages} {38005} (\bibinfo {year} {2010})}\BibitemShut
  {NoStop}%
\bibitem [{\citenamefont {Kolafa}\ \emph {et~al.}(2004)\citenamefont {Kolafa},
  \citenamefont {Lab{\'{i}}k},\ and\ \citenamefont
  {Malijevsk{\'{y}}}}]{kolafa2004}%
  \BibitemOpen
  \bibfield  {author} {\bibinfo {author} {\bibfnamefont {J.}~\bibnamefont
  {Kolafa}}, \bibinfo {author} {\bibfnamefont {S.}~\bibnamefont
  {Lab{\'{i}}k}},\ and\ \bibinfo {author} {\bibfnamefont {A.}~\bibnamefont
  {Malijevsk{\'{y}}}},\ }\href {https://doi.org/10.1039/B402792B} {\bibfield
  {journal} {\bibinfo  {journal} {Phys. Chem. Chem. Phys.}\ }\textbf {\bibinfo
  {volume} {6}},\ \bibinfo {pages} {2335} (\bibinfo {year} {2004})}\BibitemShut
  {NoStop}%
\bibitem [{\citenamefont {Henderson}\ and\ \citenamefont
  {Chan}(2000)}]{henderson2000}%
  \BibitemOpen
  \bibfield  {author} {\bibinfo {author} {\bibfnamefont {D.}~\bibnamefont
  {Henderson}}\ and\ \bibinfo {author} {\bibfnamefont {K.-Y.}\ \bibnamefont
  {Chan}},\ }\href {https://doi.org/10.1080/00268970050052051} {\bibfield
  {journal} {\bibinfo  {journal} {Mol. Phys.}\ }\textbf {\bibinfo {volume}
  {98}},\ \bibinfo {pages} {1005} (\bibinfo {year} {2000})}\BibitemShut
  {NoStop}%
\bibitem [{\citenamefont {Santos}\ \emph {et~al.}(2009)\citenamefont {Santos},
  \citenamefont {Yuste}, \citenamefont {de~Haro}, \citenamefont {Alawneh},\
  and\ \citenamefont {Henderson}}]{santos2009}%
  \BibitemOpen
  \bibfield  {author} {\bibinfo {author} {\bibfnamefont {A.}~\bibnamefont
  {Santos}}, \bibinfo {author} {\bibfnamefont {S.~B.}\ \bibnamefont {Yuste}},
  \bibinfo {author} {\bibfnamefont {M.~L.}\ \bibnamefont {de~Haro}}, \bibinfo
  {author} {\bibfnamefont {M.}~\bibnamefont {Alawneh}},\ and\ \bibinfo {author}
  {\bibfnamefont {D.}~\bibnamefont {Henderson}},\ }\href
  {https://doi.org/10.1080/00268970902852665} {\bibfield  {journal} {\bibinfo
  {journal} {Mol. Phys.}\ }\textbf {\bibinfo {volume} {107}},\ \bibinfo {pages}
  {685} (\bibinfo {year} {2009})}\BibitemShut {NoStop}%
\bibitem [{\citenamefont {Boubl{\'{i}}k}(1970)}]{boublik1970}%
  \BibitemOpen
  \bibfield  {author} {\bibinfo {author} {\bibfnamefont {T.}~\bibnamefont
  {Boubl{\'{i}}k}},\ }\href {https://doi.org/10.1063/1.1673824} {\bibfield
  {journal} {\bibinfo  {journal} {J. Chem. Phys.}\ }\textbf {\bibinfo {volume}
  {53}},\ \bibinfo {pages} {471} (\bibinfo {year} {1970})}\BibitemShut
  {NoStop}%
\bibitem [{\citenamefont {Mansoori}\ \emph {et~al.}(1971)\citenamefont
  {Mansoori}, \citenamefont {Carnahan}, \citenamefont {Starling},\ and\
  \citenamefont {Leland}}]{mansoori1971}%
  \BibitemOpen
  \bibfield  {author} {\bibinfo {author} {\bibfnamefont {G.~A.}\ \bibnamefont
  {Mansoori}}, \bibinfo {author} {\bibfnamefont {N.~F.}\ \bibnamefont
  {Carnahan}}, \bibinfo {author} {\bibfnamefont {K.~E.}\ \bibnamefont
  {Starling}},\ and\ \bibinfo {author} {\bibfnamefont {T.~W.}\ \bibnamefont
  {Leland}},\ }\href {https://doi.org/10.1063/1.1675048} {\bibfield  {journal}
  {\bibinfo  {journal} {J. Chem. Phys.}\ }\textbf {\bibinfo {volume} {54}},\
  \bibinfo {pages} {1523} (\bibinfo {year} {1971})}\BibitemShut {NoStop}%
\bibitem [{\citenamefont {Roth}\ \emph {et~al.}(2000)\citenamefont {Roth},
  \citenamefont {Evans},\ and\ \citenamefont {Dietrich}}]{roth2000}%
  \BibitemOpen
  \bibfield  {author} {\bibinfo {author} {\bibfnamefont {R.}~\bibnamefont
  {Roth}}, \bibinfo {author} {\bibfnamefont {R.}~\bibnamefont {Evans}},\ and\
  \bibinfo {author} {\bibfnamefont {S.}~\bibnamefont {Dietrich}},\ }\href
  {https://doi.org/10.1103/PhysRevE.62.5360} {\bibfield  {journal} {\bibinfo
  {journal} {Phys. Rev. E}\ }\textbf {\bibinfo {volume} {62}},\ \bibinfo
  {pages} {5360} (\bibinfo {year} {2000})}\BibitemShut {NoStop}%
\end{thebibliography}%

\end{document}